\documentclass[10pt,twocolumn,twoside]{IEEEtran}

\usepackage{subfigure}
\usepackage{citesort}
\usepackage[fleqn]{amsmath}
\usepackage{cite}
\usepackage[dvips]{graphicx}
\usepackage{amsmath}
\usepackage{amsbsy}
\usepackage{amssymb}
\usepackage{latexsym}
\newtheorem{proposition}{Proposition}

\newtheorem{theorem}{Theorem}
\newtheorem{lemma}{Lemma}

\newtheorem{remark}{Remark}

\newcommand{\migip}{\hspace*{\fill} $\Box $}

\newcommand{\euclidspace}{{\mathcal{H}}}
\newcommand{\signal}[1]{{\boldsymbol{#1}}}
\newcommand{\Natural}{{\mathbb N}}
\newcommand{\norm}[1]{\left\|#1\right\|}

\newcommand{\real}{{\mathbb R}}

\newcommand{\innerprod}[2]{\left\langle{#1},{#2}\right\rangle}

\newcommand{\refeq}[1]{(\ref{#1})}
\newcommand{\sinq}[1]{`#1'}

\newcommand{\Fix}[1]{{\rm Fix}\left({#1}\right)}

\begin{document}
%
\title{Robust Reduced-Rank Adaptive Processing
Based on Parallel Subgradient Projection and Krylov Subspace
Techniques}
%
%

 \author{Masahiro Yukawa,~\IEEEmembership{Member,~IEEE,}
Rodrigo C.~de Lamare,~\IEEEmembership{Member,~IEEE},\\
         and Isao Yamada,~\IEEEmembership{Senior Member, IEEE}
\thanks{Masahiro Yukawa is with the Amari Research Unit, RIKEN, Japan
(e-mail: myukawa@riken.jp).
This work was partly done while he was with the Department of Electronics,
University of York, UK.
}
\thanks{Rodrigo C.~de Lamare is with the Department of Electronics,
University of York, UK (e-mail:
rcdl500@ohm.york.ac.uk).
}
\thanks{Isao Yamada is with the Department of Communications and Integrated
Systems, Tokyo Institute of Technology, Japan (e-mail:
isao@comm.ss.titech.ac.jp).}
}
\markboth{IEEE TRANSACTIONS ON SIGNAL PROCESSING,
Masahiro Yukawa \MakeLowercase{\textit{et al.}}}{Yukawa \MakeLowercase{\textit{et al.}}:
Robust Reduced Rank Adaptive Filter
Based on Parallel Subgradient Projection and Krylov Subspace}
%



\maketitle







\begin{abstract}
In this paper, we propose a novel reduced-rank adaptive filtering
algorithm by blending the idea of the Krylov subspace methods with the
set-theoretic adaptive filtering framework.
Unlike the existing Krylov-subspace-based reduced-rank methods,
the proposed algorithm tracks the optimal point in the sense of
minimizing the \sinq{true} mean square error (MSE)
in the Krylov subspace, even when the estimated statistics become
erroneous (e.g., due to sudden changes of environments).
Therefore, compared with those existing methods, the proposed algorithm
is more suited to adaptive filtering applications.
The algorithm is analyzed based on a modified version of
the adaptive projected subgradient method (APSM).
Numerical examples demonstrate that the proposed algorithm
enjoys better tracking performance than the existing methods
for the interference suppression problem in code-division
multiple-access (CDMA) systems as well as for simple system
identification problems.
\end{abstract}


%

\section{Introduction}\label{sec:intro}

Reduced-rank adaptive filtering has attracted significant attention
over several research communities including signal processing; e.g.,
\cite{tufts}-\cite{dietl}. Whereas early works were motivated by the
so-called overmodeling problem, many of the recent works were
motivated mainly by computational-constraints and slow-convergence
problems due to a large number of parameters. Specifically, a Krylov
subspace associated with the input autocorrelation matrix and the
crosscorrelation vector between input and output has been used in
several methods: Cayley-Hamilton receiver \cite{moshavi}, multistage
Wiener filter (MSWF) \cite{mswf,honig&xiao,honig&goldstein},
auxiliary-vector filtering (AVF) \cite{kansal,pados}, Powers of R
(POR) receiver \cite{honig&goldstein}, and the conjugate gradient
reduced-rank filter (CGRRF) \cite{chowdhury,Wang} (see \cite{chen}
for their connections). All of those previous studies focus on
minimizing a mean square error (MSE) within the Krylov subspace (see
\cite{dietl} for linear estimation and detection in Krylov
subspaces). However, in the erroneous case (i.e., in cases where
there is a mismatch in estimates of the autocorrelation matrix and
the cross-correlation vector), the methods minimize an
\sinq{erroneous} MSE function in the Krylov subspace. Therefore, the
solution obtained at each iteration is no longer \sinq{optimal} in
the sense of minimizing the \sinq{true} MSE within the Krylov
subspace.

In this paper, we propose an adaptive technique,
named {\it Krylov reduced-rank adaptive parallel subgradient
 projection (KRR-APSP) algorithm}, tracking directly the
\sinq{optimal} solution in the Krylov subspace.
The KRR-APSP algorithm firstly performs dimensionality reduction with an
orthonormal basis of the Krylov subspace, followed by adjustments of the
coefficients of a lower-dimensional filter based on {\it the
set-theoretic adaptive filtering framework}\footnote{A related approach
called {\it set-membership adaptive filtering}
has independently been developed, e.g., in
\cite{gollamudi98:_set_lms,guo03:_f_sm_nlms}.} \cite{ysy.sp}.
As a result, in cases where the environment changes dynamically
(which makes the estimates of the statistics erroneous), the KRR-APSP
algorithm realizes better tracking capability than the existing
Krylov-subspace-based methods
(The computational complexity is comparable to the existing methods).

The rest of the paper is organized as follows.
In Section \ref{sec:motivation}, the motivation and the problem
statement are presented, in which it is shown that, in a low-dimensional
Krylov subspace, (i) the achievable MSE is close to the minimum MSE
(MMSE) and (ii) system identification of high accuracy is possible,
provided that the condition number of the autocorrelation matrix is
close to unity.
In Section \ref{sec:proposed},
we present the proposed reduced-rank algorithm,
and discuss its tracking property and computational complexity.
The KRR-APSP algorithm (i) designs multiple closed convex sets
consistent with the recently arriving data, and
(ii) moves the filter toward the intersection of the convex sets
(to find a feasible solution) by means of parallel subgradient
projection at each iteration.
Because the noise is taken into account in the set design, KRR-APSP is
intrinsically robust.
In Section \ref{sec:analysis}, to prove important properties
({\it monotonicity} and {\it asymptotic optimality}) of the proposed algorithm,
we firstly present an alternative derivation of
the algorithm from a modified version of
{\it the adaptive projected subgradient method
(APSM)}\footnote{APSM has proven
a promising tool to derive efficient algorithms in many applications
\cite{reya04_ds_cdma,yry_j_ieice05,yuya_power_ieee,yumuya_eurasip,syy_icassp06,ysy_apqp2007,cavalcante08,slavakis_ieee08}.}
\cite{yamada03_kaisetsu,yagu_paper},
and then present an analysis of the modified APSM.
It is revealed that,
in the (original) high dimensional vector space, the proposed algorithm
performs parallel subgradient projection in a series of Krylov
subspaces.
In Section \ref{sec:numer_examp}, numerical examples are presented
to verify the advantages of the proposed algorithm over CGRRF,
followed by the conclusion in Section \ref{sec:conclusion}.

\section{Motivation and Problem Statement}\label{sec:motivation}

\begin{figure}[t]
\centering
 \includegraphics[width=6cm]{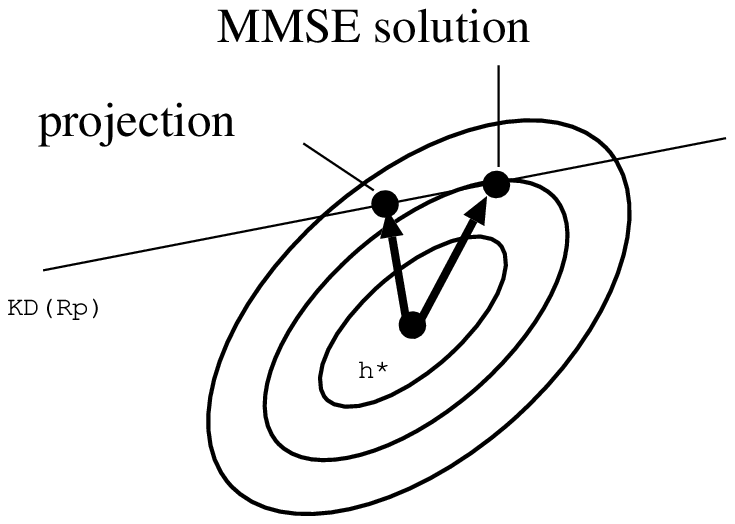}
 \caption{$P_{\mathcal{K}_D(\signal{R},\signal{p})} (\signal{h}^*)$
and $P_{\mathcal{K}_D(\signal{R},\signal{p})}^{(\signal{R})}
 (\signal{h}^*)$ with the equal error contours of the MSE surface.}
\label{fig:two_solutions}
\end{figure}

Let $\real$, $\Natural$, and $\Natural^*$ denote the sets of all real
numbers, nonnegative integers, and positive integers, respectively.
We consider the following linear model:
\begin{equation}
 d_k := \signal{u}_k^T\signal{h}^* + n_k, \ \forall k\in\Natural,\label{eq:data}
\end{equation}
where $\signal{u}_k:=[u_k,u_{k-1}, \cdots,u_{k-N+1}]^T\in\real^N$
($N\in\Natural^*$) denotes the input vector,
$\signal{h}^*\in\real^N$ the unknown system, $n_k$ the additive noise,
and $d_k$ the output ($k$: sample index, $(\cdot)^T$:
{\it transposition}).
The MMSE filter in the whole space $\real^N$ is well-known to be
characterized by the so-called Wiener-Hopf equation
$\signal{R}\signal{h}_{\rm MMSE}=\signal{p}$ (see, e.g., \cite{haykin}),
where $\signal{R}:={\rm E}\{ \signal{u}_k\signal{u}_k^T\}$ and
$\signal{p}:={\rm E}\{ \signal{u}_k d_k\}$
(${\rm E}\{\cdot\}$: {\it expectation}).
For simplicity, we assume that $\signal{R}$ is invertible and
the input and the noise are (statistically) orthogonal; i.e.,
$E\{n_k\signal{u}_k\}=\signal{0}$.
In this case, $\signal{p}={\rm E}\{ \signal{u}_k
(\signal{u}_k^T\signal{h}^* + n_k)\}= \signal{R}\signal{h}^*$, and
the MSE function $f:\real^N\rightarrow [0,\infty)$ is given as
\begin{align}
\hspace*{-1.5em} f(\signal{h}):=&
{\rm E}\{ (d_k - \signal{h}^T \signal{u}_k)^2 \}
=\signal{h}^T\signal{R}\signal{h} - 2\signal{h}^T\signal{p} +
 \sigma_d^2\nonumber\\
=&\norm{\signal{h} - \signal{h}^*}_{\signal{R}}^2 -
 \norm{\signal{h}^*}_{\signal{R}}^2
+\sigma_d^2.\label{eq:MSE}
\end{align}
Here, $\sigma_d^2:={\rm E}\{d_k^2\}$ and
$\norm{\cdot}_{\signal{R}}$ is the $\signal{R}$-norm\footnote{
The $\signal{R}$-norm is also called {\it the energy norm induced by
$\signal{R}$}.
The same norm is used in \cite{dietl_SPIE01} to derive the CG method.}
defined for any vector $\signal{a}\in\real^N$ as
$\norm{\cdot}_{\signal{R}}:=\sqrt{\signal{a}^T\signal{R}\signal{a}}$.
From \refeq{eq:MSE}, it is seen that
$\signal{h}^* =\signal{h}_{\rm MMSE}(=\signal{R}^{-1}\signal{p})$.

Let us now consider, for $D\in\{1,2,\cdots,N\}$, the MMSE filter
within the following Krylov subspace:
\begin{align}
\hspace*{-1em} \mathcal{K}_D(\signal{R},\signal{p}) :=& {\rm span}
  \{\signal{p},\signal{R}\signal{p},\cdots, \signal{R}^{D-1}\signal{p}\}\\
=& {\rm span}
  \{\signal{R}\signal{h}^*,\signal{R}^2\signal{h}^*,\cdots, \signal{R}^D\signal{h}^*\}
\subset \real^N.
\end{align}
Referring to \refeq{eq:MSE}, the MMSE solution in
$\mathcal{K}_D(\signal{R},\signal{p})$
is characterized by
\begin{equation}
 P_{\mathcal{K}_D(\signal{R},\signal{p})}^{(\signal{R})} (\signal{h}^*) \in
\arg\min_{\signal{h}\in \mathcal{K}_D(\signal{R},\signal{p})}
\norm{\signal{h}^* - \signal{h}}_{\signal{R}}, \label{eq:best_approximation_in_Rnorm}
\end{equation}
where we denote by $P_C^{(\signal{A})}(\signal{x})$
the metric projection of a vector $\signal{x}$ onto a closed convex set $C$
in the $\signal{A}$-norm sense.
In particular, the metric projection in the sense of Euclidean norm is
denoted simply by $P_C(\signal{x})$.
In words, the MMSE filter in the subspace is the best approximation, in
the $\signal{R}$-norm sense, of $\signal{h}^*$ in
$\mathcal{K}_D(\signal{R},\signal{p})$.
Noting that $P_{\mathcal{K}_D(\signal{R},\signal{p})}^{(\signal{R})}
(\signal{h}^*)$ coincides with the vector obtained through $D$ steps
of the conjugate gradient (CG) method with its initial point being
the zero vector, the MSE is bounded as follows
\cite[Theorem 10.2.6]{golub}:
\begin{equation}
 f(P_{\mathcal{K}_D(\signal{R},\signal{p})}^{(\signal{R})}
  (\signal{h}^*))
\leq
\left[4
\left(\frac{\sqrt{\kappa}-1}{\sqrt{\kappa}+1}
\right)^{2D}
-1 \right]
\norm{\signal{h}^*}_{\signal{R}}^2
+\sigma_d^2,
\end{equation}
where $\kappa:=\norm{\signal{R}}_2 \norm{\signal{R}^{-1}}_2\geq 1$ is
the condition number of $\signal{R}$.
System identifiability in $\mathcal{K}_D(\signal{R},\signal{p})$ is
discussed below.
\begin{remark}
How accurately can the system $\signal{h}^*$ be identified in the subspace
$\mathcal{K}_D(\signal{R},\signal{p})$?
In the system identification problem, we wish to minimize the Euclidean
norm $\norm{\signal{h}^*  - \signal{h}}$ rather than the $\signal{R}$-norm
$\norm{\signal{h}^*  - \signal{h}}_{\signal{R}}$.
To clarify the difference between the MSE minimization and the system
identification over $\mathcal{K}_D(\signal{R},\signal{p})$,
the projections in the different senses are illustrated in Fig.~\ref{fig:two_solutions}.
By the Rayleigh-Ritz theorem \cite{horn85},
it is readily verified that
$\lambda_{\max}^{-1/2}\norm{\signal{x}}_{\signal{R}}
\leq \norm{\signal{x}}\leq
\lambda_{\min}^{-1/2}\norm{\signal{x}}_{\signal{R}}$ for any
 $\signal{x}\in\real^N$, where $\lambda_{\max}>0$ and $\lambda_{\min}>0$
 denote the maximum and minimum eigenvalues of $\signal{R}$,
 respectively.
It is thus verified that
$\norm{P_{\mathcal{K}_D(\signal{R},\signal{p})}(\signal{h}^*)
-
P_{\mathcal{K}_D(\signal{R},\signal{p})}^{(\signal{R})}(\signal{h}^*)}
\leq
\norm{\signal{h}^* - P_{\mathcal{K}_D(\signal{R},\signal{p})}^{(\signal{R})}
(\signal{h}^*)}
\leq
\lambda_{\min}^{-1/2}\norm{\signal{h}^* -
 P_{\mathcal{K}_D(\signal{R},\signal{p})}^{(\signal{R})}
(\signal{h}^*)}_{\signal{R}}
\leq 2\lambda_{\min}^{-1/2}
\norm{\signal{h}^*}_{\signal{R}}\alpha^D(\kappa)$,
where  $\alpha(\kappa):=(\sqrt{\kappa}-1)/(\sqrt{\kappa}+1)\in[0,1)$.
Here, the first inequality is due to the basic property of projection,
 and the third one is verified by \cite[Theorem 10.2.6]{golub}.
This suggests that system identification of high accuracy would be
possible for a small $D$ when $\kappa\approx 1$ (If $\kappa\gg 1$,
preconditioning\footnote{The importance of preconditioning is
well-known in numerical linear algebra; see, e.g.,
\cite{axelsson_BIT85,saad_book03} and the references therein.
Also the importance is mentioned in \cite{hull91} for an application of
the conjugate gradient method to the adaptive filtering problem.
Different types of CG-based adaptive filtering algorithms have also been
 proposed, e.g., in \cite{boray92,chang00}.}
should be performed).
\migip
\end{remark}

\begin{figure}[t]
\centering
 \includegraphics[width=8cm]{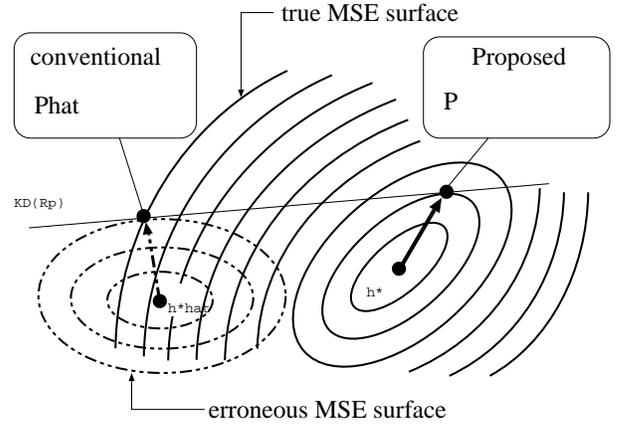}
 \caption{An illustration of the goal of this paper.
\sinq{Conventional} stands for the conventional Krylov-subspace-based
 methods such as CGRRF.}
\label{fig:geom_vr}
\end{figure}


In reality, $\signal{R}$ and $\signal{p}$ are rarely available,
thus should be estimated from observed measurements.
Let $\widehat{\signal{R}}$ and $\widehat{\signal{p}}$ be
estimates of $\signal{R}$ and $\signal{p}$, respectively, and
$\widehat{\signal{h}}^*$ be characterized by
$\widehat{\signal{R}}\widehat{\signal{h}}^*=\widehat{\signal{p}}$.
CGRRF \cite{chowdhury01,dietl_VTC01,burykh_EUSIPCO02}
computes, at each iteration,
the best approximation of $\widehat{\signal{h}}^*$
in $\mathcal{K}_D(\widehat{\signal{R}},\widehat{\signal{p}})$
in the $\widehat{\signal{R}}$-norm sense; i.e.,
$P_{\mathcal{K}_D(\widehat{\signal{R}},\widehat{\signal{p}})}^{(\widehat{\signal{R}})}(\widehat{\signal{h}}^*)$.
This realizes significantly fast convergence and reasonable steady-state
performance as long as good estimates are available; i.e.,
$\widehat{\signal{R}}\approx\signal{R}$ and
$\widehat{\signal{p}}\approx\signal{p}$.
However, once those estimates become unreliable (which happens when
the environments change suddenly),
$P_{\mathcal{K}_D(\widehat{\signal{R}},\widehat{\signal{p}})}^{(\widehat{\signal{R}})}(\widehat{\signal{h}}^*)$
makes little sense, and CGRRF (or the other existing
Krylov-subspace-based methods) should wait until a certain amount of
data arrive to recapture reasonable estimates.

The goal of this paper is to propose an alternative to the existing
Krylov-subspace-based methods to address this restriction.
To be specific, the main problem in this work is stated as follows.
Given that the Krylov subspace is employed for dimensionality reduction,
the problem is to design an efficient algorithm that can always track
$P_{\mathcal{K}_D(\widehat{\signal{R}},\widehat{\signal{p}})}^{(\signal{R})}(\signal{h}^*)$,
which minimizes the true MSE $f(\signal{h})$ over
$\mathcal{K}_D(\widehat{\signal{R}},\widehat{\signal{p}})$ [see
\refeq{eq:MSE}].
Such an algorithm should have better tracking capability
than the existing methods after dynamic changes of environments,
because
$P_{\mathcal{K}_D(\widehat{\signal{R}},\widehat{\signal{p}})}^{(\widehat{\signal{R}})}(\widehat{\signal{h}}^*)$
does not minimize the true MSE as long as the estimates
$\widehat{\signal{R}}$ and $\widehat{\signal{p}}$ are erroneous.
The concept is illustrated in Fig.~\ref{fig:geom_vr}, in which the
estimates are assumed to become erroneous.
Note in the figure that the difference between $f(\signal{h})$ and
$\norm{\signal{h} - \signal{h}^*}_{\signal{R}}^2$ is a constant in terms
of $\signal{h}$, which makes no difference in the equal error contours.
In the following section,
we present an adaptive algorithm that achieves the goal.

\section{Proposed Reduced-Rank Adaptive Filter}\label{sec:proposed}

We firstly present a reduced-rank version of the set-theoretic adaptive
filtering algorithm named {\it adaptive parallel subgradient projection
(APSP) algorithm} \cite{ysy.sp}.
The proposed algorithm is called {\it Krylov Reduced-Rank Adaptive
Parallel Subgradient Projection (KRR-APSP)}.
We then show, for its simplest case, that the proposed algorithm tracks
$P_{\mathcal{K}_D(\widehat{\signal{R}},\widehat{\signal{p}})}^{(\signal{R})}(\signal{h}^*)$,
and discuss its computational complexity.

\subsection{Proposed KRR-APSP Algorithm}\label{subsec:strraf}

\begin{figure}[t]
\centering
 \includegraphics[width=8cm]{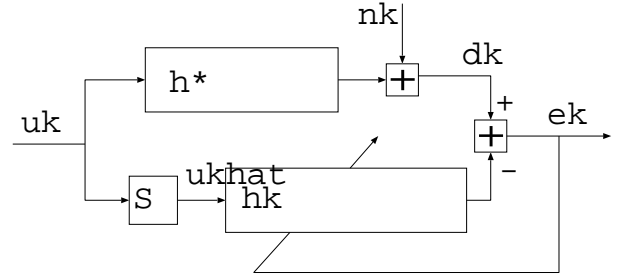}
 \caption{Reduced-rank adaptive filtering scheme.}
\label{fig:rrfilter}
\end{figure}

Let $\widehat{\signal{R}}_k$ and $\widehat{\signal{p}}_k$ be estimates
of $\signal{R}$ and $\signal{p}$ at time $k\in\Natural$, respectively,
and $\signal{S}_k$ an $N\times D$ matrix whose column vectors form
an orthonormal basis\footnote{The orthonormality is essential in the
analysis (see Section \ref{subsec:analysis}).}
(in the sense of the standard inner product) of the subspace
$\mathcal{K}_D(\widehat{\signal{R}}_k,\widehat{\signal{p}}_k)$.
For dimensionality reduction, we force the adaptive filter
$\signal{h}_k\in\real^N$ to lie in
$\mathcal{K}_D(\widehat{\signal{R}}_k,\widehat{\signal{p}}_k)\subset\real^N$
at each time instance $k$.
Thus, with a lower dimensional vector $\widetilde{\signal{h}}_k\in\real^D$,
the adaptive filter is characterized as
$\signal{h}_k=\signal{S}_k\widetilde{\signal{h}}_k$.
In the following, a tilde will be used for expressing
a $D$-dimensional vector (or a subset of $\real^D$).
The output of the adaptive filter is given by
\begin{equation}
 \signal{h}_k^T\signal{u}_k=\widetilde{\signal{h}}_k^T
\signal{S}_k^T\signal{u}_k=
\widetilde{\signal{h}}_k^T\widetilde{\signal{u}}_k \quad
(\widetilde{\signal{u}}_k:=\signal{S}_k^T\signal{u}_k\in\real^D).
\end{equation}
The reduced-rank adaptive filtering scheme is illustrated
in Fig.~\ref{fig:rrfilter}.

The idea of set-theoretic adaptive filtering is as follows:
\begin{enumerate}
 \item construct (possibly multiple) closed convex sets containing
a desired filter,
i.e.~$P_{\mathcal{K}_D(\widehat{\signal{R}}_k,\widehat{\signal{p}}_k)}^{(\signal{R})}(\signal{h}^*)$
in this case, with high probability; and
 \item approach the intersection of those sets at each iteration.
\end{enumerate}
Let us present the design of the closed convex sets.
Given $r\in\Natural^*$, we define
\begin{eqnarray*}
\signal{U}_k &:=& \left[
\signal{u}_k, \signal{u}_{k-1}, \cdots,
\signal{u}_{k-r+1}\right] \in\real^{N\times r}\\
\signal{d}_k &:=& \left[
d_k, d_{k-1}, \cdots,
d_{k-r+1}\right] \in\real^r\\
\signal{e}_k (\signal{h})
&:=& \signal{U}_k^T\signal{h} - \signal{d}_k \in\real^r, \ \forall \signal{h}\in\real^N.
\end{eqnarray*}
Then, with a simple restriction on $\signal{h}\in\real^N$ in the
stochastic property set proposed in \cite{ysy.sp}, the closed convex sets in
$\real^N$ are given as
\begin{align}
\hspace*{-1.8em} C_k(\rho):=\left\{\signal{h}\in\mathcal{R}(\signal{S}_k):
         g_k(\signal{h}):= \norm{\signal{e}_k(\signal{h})}^2 -\rho\leq
         0 \right\}, &\nonumber\\
 k\in\Natural,&\label{eq:stochastic}
\end{align}
where $\rho\geq 0$,
$\mathcal{R}(\cdot)$ stands for {\it range},  and $\norm{\cdot}$
denotes the Euclidean norm.
Intuitively, $C_k(\rho)$ is a set of filtering vectors consistent with
the data observed at time $k$ in the sense that the norm of the
error-vector is bounded by a small constant $\rho$.
If $\rho$ is too small, there could be no consistent solution;
for an extreme example, if $\rho=0$ and we have the data sets
$(\signal{u}_{k_1},d_{k_1})$ and $(\signal{u}_{k_2},d_{k_2})$ such that
$\signal{u}_{k_1}=\signal{u}_{k_2}$ and $d_{k_1}\neq d_{k_2}$
($k_1$, $k_2\in\Natural$), then
$C_{k_1}(\rho)\cap C_{k_2}(\rho)=\emptyset$.
Note however that, even in such an infeasible case, the proposed
algorithm is guaranteed to move the filter closer to all the points
that minimize a weighted sum of the distances to the convex sets
$(C_k(\rho))_{k\in\Natural}$,
as will be shown in Theorem \ref{prop:optimality}.a in Section
\ref{subsec:analysis}.
The design of $\rho$ is involved with the noise
statistics (see \cite{ysy.sp}).


Let $\mathcal{I}_k$ be the control sequence at the $k$th iteration;
i.e., the set of indices used at time $k$ (a typical example is
$\mathcal{I}_k:=\{k,k-1,\cdots,k-q+1\}$ for $q\in\Natural^*$).
Replacing $\signal{h}$ in $C_\iota(\rho)$, $\iota\in\mathcal{I}_k$, by
$\signal{S}_k\widetilde{\signal{h}}$, the stochastic property set
in $\real^D$ is obtained as follows:
\begin{align}
 &\hspace*{-1em}\widetilde{C}_\iota^{(k)}(\rho):=\left\{\widetilde{\signal{h}}\in\real^D:
         g_\iota^{(k)}(\widetilde{\signal{h}}):=
\norm{\signal{e}_\iota^{(k)}(\widetilde{\signal{h}})}^2 -\rho\leq
         0 \right\},\nonumber\\
&\hspace*{14.5em}\iota\in\mathcal{I}_k, \ k\in\Natural.
\label{eq:stochastic_lowerdim}
\end{align}
Here,
$\signal{e}_\iota^{(k)} (\widetilde{\signal{h}}):=
\signal{U}_\iota^T\signal{S}_k\widetilde{\signal{h}} - \signal{d}_\iota \in\real^r,
 \ \forall \widetilde{\signal{h}}\in\real^D$.
The projection onto $\widetilde{C}_\iota^{(k)}(\rho)$ is approximated
by the projection onto the simple closed half-space
$\widetilde{H}_{\iota,k}^-(\widetilde{\signal{h}}_k)\supset
\widetilde{C}_\iota^{(k)}(\rho)$ defined as
\begin{eqnarray}
\hspace*{0em}\widetilde{H}_{\iota,k}^-(\widetilde{\signal{h}}_k):=
\left\{
\widetilde{\signal{h}}\in\real^D:
\innerprod{\widetilde{\signal{h}} - \widetilde{\signal{h}}_k}{
\widetilde{\signal{s}}_\iota^{(k)}}
 +
g_\iota^{(k)}(\widetilde{\signal{h}}_k)\leq 0
\right\},\nonumber\\
\hspace*{-5em}\iota\in\mathcal{I}_k, \ k\in\Natural.\hspace*{3em}
\label{eq:Htilde-}
\end{eqnarray}
where
$\widetilde{\signal{s}}_\iota^{(k)}:=\signal{\nabla}g_\iota^{(k)}(\widetilde{\signal{h}}_k):=
2\signal{S}_k^T\signal{U}_\iota\signal{e}_\iota^{(k)}
(\widetilde{\signal{h}}_k)\in\real^D$.
An important property is
$\widetilde{\signal{h}}_k\not\in \widetilde{C}_\iota^{(k)}(\rho) \Rightarrow
\widetilde{\signal{h}}_k\not\in \widetilde{H}_{\iota,k}^-(\widetilde{\signal{h}}_k)$ \cite[Lemma 2]{ysy.sp},
thus the boundary of $\widetilde{H}_{\iota,k}^-(\widetilde{\signal{h}}_k)$ is a separating
hyperplane between $\widetilde{\signal{h}}_k$ and $\widetilde{C}_\iota^{(k)}(\rho)$.
The projection of $\widetilde{\signal{h}}_k$ onto
$\widetilde{H}_{\iota,k}^-(\widetilde{\signal{h}}_k)$ is given as
\begin{equation}
P_{\widetilde{H}_{\iota,k}^-(\widetilde{\signal{h}}_k)}(\widetilde{\signal{h}}_k)
= \left\{
\begin{array}{ll}
\widetilde{\signal{h}}_k &  \mbox{if }
 g_\iota^{(k)}(\widetilde{\signal{h}}_k)
 \leq 0,\\
\widetilde{\signal{h}}_k -
 \displaystyle\frac{g_\iota^{(k)}(\widetilde{\signal{h}}_k)}
{\norm{\widetilde{\signal{s}}_\iota^{(k)}}^2}
\widetilde{\signal{s}}_\iota^{(k)}&
 \mbox{otherwise},
\end{array}
\right.\label{eq:proj_lower}
\end{equation}
which is also referred to as the {\it subgradient projection\footnote{
Although the function $g_\iota^{(k)}$ is differentiable,
the subgradient projection can be defined also for non-differentiable
functions.
Note that lev$_{\leq 0}
g_\iota^{(k)}:=\{\widetilde{\signal{h}}\in\real^D:g_\iota^{(k)}(\widetilde{\signal{h}})\leq
0\}\neq \emptyset$.
}
relative to}
$g_\iota^{(k)}$ (see Appendix \ref{append:mathdef}).
Let $w_\iota^{(k)}\in(0,1]$, $\iota\in\mathcal{I}_k$, $k\in\Natural$,
denote the weight satisfying $\sum_{\iota\in\mathcal{I}_k} w_\iota^{(k)}
=1$; see \cite{yuya_power_ieee}
for a strategic design of the weights.
Then, the proposed KRR-APSP algorithm is presented in what follows.

Given an arbitrary initial vector $\widetilde{\signal{h}}_0\in\real^D$,
the sequence $(\widetilde{\signal{h}}_k)_{k\in\Natural}\subset\real^D$
is inductively generated as follows.
Given $\signal{h}_k$ and $\mathcal{I}_k$ at each time $k\in\Natural$,
$\signal{h}_{k+1}$ is defined as
\begin{equation}
\hspace*{-2em}\widetilde{\signal{h}}_{k+1} =
\widetilde{\signal{h}}_k + \lambda_k
\mathcal{M}_k \left(\displaystyle\sum_{\iota\in\mathcal{I}_k}
w_\iota^{(k)}
P_{\widetilde{H}_{\iota,k}^-(\widetilde{\signal{h}}_k)}(\widetilde{\signal{h}}_k)
- \widetilde{\signal{h}}_k \right),\label{eq:update_prop_low}
\end{equation}
where $\lambda_k\in[0,2]$,
$\widetilde{H}_{\iota,k}^-(\widetilde{\signal{h}}_k)$ is defined as in \refeq{eq:Htilde-},
 and
\begin{align*}
&\hspace*{-2em}\mathcal{M}_k:= \\
&\hspace*{-2em}\left\{
\begin{array}{ll}
1 &  \mbox{if }
 g_\iota^{(k)}(\widetilde{\signal{h}}_k)
 \leq 0, \ \forall \iota\in\mathcal{I}_k, \\
\displaystyle\frac{\displaystyle\sum_{\iota\in\mathcal{I}_k}
w_\iota^{(k)}
\norm{P_{\widetilde{H}_{\iota,k}^-(\widetilde{\signal{h}}_k)}(\widetilde{\signal{h}}_k)
- \widetilde{\signal{h}}_k}^2}
{\norm{\displaystyle\sum_{\iota\in\mathcal{I}_k}
w_\iota^{(k)}
P_{\widetilde{H}_{\iota,k}^-(\widetilde{\signal{h}}_k)}(\widetilde{\signal{h}}_k)
- \widetilde{\signal{h}}_k}^2}&
 \mbox{otherwise}.
\end{array}
\right.
\end{align*}

\begin{table}
\caption{Efficient Implementation of the Proposed Algorithm.}\label{table:rrank_psp}
\begin{tabular}{l}
\hline\\[-.5em]
Requirements: Initial transformation matrix $\signal{S}_0$,
inputs $(\signal{U}_k)_{k\in\Natural}$, \\
outputs $(\signal{d}_k)_{k\in\Natural}$,
control sequence $\mathcal{I}_k$,
step size $\lambda_k\in[0,2]$,\\
weights $w_\iota^{(k)}$, $\forall \iota\in\mathcal{I}_k$,
initial vector $\widetilde{\signal{h}}_0\in\real^D$,
constant $\rho\geq 0$, $m\in\Natural^*$
\end{tabular}
\begin{tabular}{ll}
1. & Filter output: $y_k:=
 \widetilde{\signal{u}}_k^T\widetilde{\signal{h}}_k
(=\signal{u}_k^T\signal{S}_k\widetilde{\signal{h}}_k)$\\
2. & Filter update: \\
 &(a)  {\bf For} $\iota\in\mathcal{I}_k$, do the following: \\
& \hspace*{1.1em}
$\signal{U}_\iota^{(k)}:=\signal{S}_k^T\signal{U}_\iota\in
\real^{D\times r}$ \\[.3em]
& \hspace*{1.5em} $\signal{e}_\iota^{(k)} :=
(\signal{U}_\iota^{(k)})^T
\widetilde{\signal{h}}_k - \signal{d}_\iota\in\real^r$
\\
& \hspace*{2.5em} $\mbox{{\bf If} } \norm{\signal{e}_\iota^{(k)}}^2
 \leq \rho$,\\
& \hspace*{3.5em} $\widetilde{\signal{\delta}}_\iota^{(k)} :=\signal{0}\in\real^D$,
 $\ell_\iota^{(k)}:=0$\\
& \hspace*{2.5em} {\bf else}\\
& \hspace*{3.5em} $\signal{a}_\iota^{(k)}:=\signal{U}_\iota^{(k)} \signal{e}_\iota^{(k)}\in\real^D$\\
& \hspace*{3.5em} $c_\iota^{(k)}:=\norm{\signal{a}_\iota^{(k)}}^2\in[0,\infty)$\\
& \hspace*{3.5em} $d_\iota^{(k)}:=\rho -\norm{\signal{e}_\iota^{(k)}}^2\in(-\infty,\rho] $\\
& \hspace*{3.5em} $\widetilde{\signal{\delta}}_\iota^{(k)} :=
w_\iota^{(k)} d_\iota^{(k)} \signal{a}_\iota^{(k)}/(2c_\iota^{(k)})\in\real^D$\\[-.5em]
& \\
& \hspace*{3.5em} $\ell_\iota^{(k)}:=
\left(\norm{\signal{\delta}_\iota^{(k)}}^2/w_\iota^{(k)}=\right)
w_\iota^{(k)} (d_\iota^{(k)})^2/(4
 c_\iota^{(k)}) \in (0,\infty)$\\
& \hspace*{2.5em} {\bf endif;}\\
& \hspace*{1.5em}{\bf end;}\\
\\[-.5em]
& (b) {\bf If} $\norm{\signal{e}_\iota^{(k)}}^2 \leq \rho$
for all $\iota\in\mathcal{I}_k$,\\
& \\
& \hspace*{2.5em} $\widetilde{\signal{h}}_{k+1} :=\widetilde{\signal{h}}_k\in\real^D$\\
& \hspace*{1.2em} {\bf else}\\
& \hspace*{2.5em} $\widetilde{\signal{f}}_k:= \displaystyle\sum_{\iota\in\mathcal{I}_k}
\widetilde{\signal{\delta}}_\iota^{(k)}\in\real^D$ \\
&\hspace*{2.5em}  $\mathcal{M}_k :=
\norm{\widetilde{\signal{f}}_k}^{-2}
\displaystyle\sum_{\iota\in\mathcal{I}_k}\ell_\iota^{(k)}
\in[1,\infty)$
\\
&\hspace*{2.5em}  $\widetilde{\signal{h}}_{k+1} :=\widetilde{\signal{h}}_k +
\lambda_k\mathcal{M}_k \widetilde{\signal{f}}_k\in\real^D$\\
& \hspace*{1.5em}{\bf endif;}\\
3: &{\bf if} $k\equiv 1$ mod $m$\\
& \hspace*{1em}Compute $\signal{S}_{k+1}\in\real^{N\times D}$, an
 orthonormalized version of\\
& \hspace*{1em}$\signal{K}_D(\widehat{\signal{R}}_k,\widehat{\signal{p}}_k)$; see
 Section \ref{subsec:complexity}\\
& {\bf else}\\
& \hspace*{1em}$\signal{S}_{k+1}:=\signal{S}_k$\\
& {\bf endif;}\\
\hline
\end{tabular}
\end{table}

For convenience, efficient implementation of
the proposed algorithm is given in TABLE \ref{table:rrank_psp}
(For computational efficiency, we introduce a parameter $m$ to
control how frequently $\signal{S}_k$ is updated).
We mention that, although the condition for updating
$\widetilde{\signal{\delta}}_\iota^{(k)}$ is similar to the one used in
{\it the set-membership affine projection algorithm} \cite{werner01},
the major differences are that
(i) the update is based on the subgradient projection,
(ii) multiple closed convex sets are employed at each iteration
(each set is indicated by an element of $\mathcal{I}_k$), and
(iii) no matrix inversion is required.

We shall finish up this subsection by summarizing the parameters used in
the proposed algorithm:
\begin{itemize}
 \item $r$: the dimension of the orthogonal complement of the underlying
       subspace of $C_k(0)$ (see the definition of $\signal{U}_k$, and
       $\signal{d}_k$ before \refeq{eq:stochastic}),
 \item $q$: the number of projections computed at each iteration,
 \item $\rho$: the error bound (controlling the \sinq{volume} of
       $C_k(\rho)$),
 \item $m$: the frequency of updating $\signal{S}_k$.
\end{itemize}
Intuitively, the convex set $C_k(\rho)$ is obtained by
\sinq{ballooning} the linear variety used in the affine projection
algorithm (APA) \cite{hinamoto,ozeki}, and $r$ corresponds to the
\sinq{order} of APA \cite{haykin}.

The tracking property and the computational complexity of the proposed
algorithm are discussed in the following subsection.

\subsection{Tracking Property and Computational Complexity}\label{subsec:complexity}
As explained in the final paragraph in Section \ref{sec:motivation},
an algorithm that tracks
$P_{\mathcal{K}_D(\widehat{\signal{R}}_k,\widehat{\signal{p}}_k)}^{(\signal{R})}(\signal{h}^*)$
is expected to enjoy better tracking capability than the existing
Krylov-subspace-based reduced-rank methods.
In this subsection, we firstly show that the proposed algorithm
(or the vector $\signal{h}_k(=\signal{S}_k\widetilde{\signal{h}}_k)$,
$k\in\Natural$,
generated by the proposed algorithm) has such a property
for its simplest case: $r=1$, $\rho=0$, $\mathcal{I}_k=\{k\}$ (i.e.,
$q=1$).
In this case, the proposed algorithm is reduced to
\begin{equation}
 \widetilde{\signal{h}}_{k+1} =  \widetilde{\signal{h}}_k +
  \bar{\lambda}_k\frac{d_k - \widetilde{\signal{h}}_k^T
  \widetilde{\signal{u}}_k}{\norm{\widetilde{\signal{u}}_k}^2}\widetilde{\signal{u}}_k,\label{eq:simplest}
\end{equation}
where $\bar{\lambda}_k:=\lambda_k/2\in[0,1]$.
The update equation in \refeq{eq:simplest} is nothing but the NLMS
algorithm
(It should be mentioned that the step-size range of $\bar{\lambda}_k$ is
a half of that of NLMS).
Thus, \refeq{eq:simplest} is a stochastic gradient algorithm for the
following problem:
\begin{equation}
 \min_{\widetilde{\signal{h}}\in\real^D}
{\rm E}\{ (d_k - \widetilde{\signal{h}}^T \widetilde{\signal{u}}_k)^2
\}. \label{eq:prob_equivalent}
\end{equation}
This implies that $\widetilde{\signal{h}}_k$ generated by \refeq{eq:simplest}
tracks the minimizer of \refeq{eq:prob_equivalent}; for details about
the tracking performance of NLMS, see \cite{sayed_book03} and the
references therein.
Hence, noting that
$\widetilde{\signal{u}}_k=\signal{S}_k^T\signal{u}_k$, it is seen that
$\signal{h}_k(:=\signal{S}_k\widetilde{\signal{h}}_k)$ tracks
the solution to the following problem (which is equivalent to
\refeq{eq:prob_equivalent}):
\begin{equation}
 \min_{\signal{h}\in\mathcal{R}(\signal{S}_k)}
{\rm E}\{ (d_k - \signal{h}^T \signal{u}_k)^2
\}. \label{eq:problem_projection}
\end{equation}
Referring to \refeq{eq:MSE} and \refeq{eq:best_approximation_in_Rnorm},
the minimizer of \refeq{eq:problem_projection} is
$P_{\mathcal{K}_D(\widehat{\signal{R}}_k,\widehat{\signal{p}}_k)}^{(\signal{R})}(\signal{h}^*)$.
This verifies that $\signal{h}_k(=\signal{S}_k\widetilde{\signal{h}}_k)$
generated by \refeq{eq:simplest} tracks
$P_{\mathcal{K}_D(\widehat{\signal{R}}_k,\widehat{\signal{p}}_k)}^{(\signal{R})}(\signal{h}^*)$.

Now, let us move to the discussion about the computational complexity
(i.e., the number of multiplications per iteration) of the proposed
algorithm.
For simplicity, we let $\mathcal{I}_k:=\{k,k-1,\cdots,k-q+1\}$,
which is used in Section \ref{sec:numer_examp}.
We assume that,
given $\widehat{\signal{R}}_k$ and $\widehat{\signal{p}}_k$,
the complexity to construct the matrix $\signal{S}_k$
is the same as that of CGRRF\footnote{The Lanczos method, which is
essentially equivalent to the CG method \cite{golub}, can also be used
for constructing $\signal{S}_k$.}.
As $\signal{S}_k$ is computed every $m$ iterations (see TABLE
\ref{table:rrank_psp}), the average complexity for computing
$\signal{S}_k$ is $(D-1)N^2/m + (5D-4)N/m +2(D-1)/m$.

What about the complexity to update
$\widehat{\signal{R}}_k$ and $\widehat{\signal{p}}_k$?
For the system model presented in Section \ref{sec:motivation},
the autocorrelation matrix $\signal{R}$ is known to have
{\it a Toeplitz structure}, provided that the input process is
stationary.
Hence, it is sufficient to estimate $E\{u_k\signal{u}_k\}\in\real^N$,
which can be done by\footnote{If, for example, the system model
presented in Section \ref{subsec:cdma_simulations} is to be considered, then
$\signal{R}$ is {\it not} Toeplitz in general.
In such a case, at least the upper triangular portion of $\signal{R}$
should be estimated (Note that $\signal{R}$ is always symmetric).}
$\widehat{\signal{r}}_{k+1}:=\gamma\widehat{\signal{r}}_k +
u_k\signal{u}_k$, $k\in\Natural$, with the forgetting factor
$\gamma\in(0,1)$.
Similarly, the vector $\widehat{\signal{p}}_k$ is updated as
$\widehat{\signal{p}}_{k+1}:=\gamma\widehat{\signal{p}}_k +
d_k\signal{u}_k$, $k\in\Natural$.
Thus, the complexity for updating $\widehat{\signal{R}}_k$
and $\widehat{\signal{p}}_k$ is $4N$.

The rest is the complexity for the filter update.
One of the distinguished advantages of the APSP algorithm is its
{\it inherently parallel structure}
\cite{combettes_foundations,bauschke.borwein,censor.book,butnariu2001,ysy.sp,yuya_power_ieee}.
We start by considering the case where only a single processor is
available.
Because the matrices $(\signal{U}_\iota)_{\iota\in\mathcal{I}_k}$,
used at time $k$, have only $q+r-1$ distinct column vectors
($\signal{u}_k$, $\signal{u}_{k-1}$, $\cdots$,$\signal{u}_{k-q-r+2}$),
the complexity to compute $\signal{U}_\iota^{(k)}$ for all
$\iota\in\mathcal{I}_k$ is $(q+r-1)DN$.
Fortunately, however, this is only required when $\signal{S}_k$ is
updated (every $m$ iterations), and,
when $\signal{S}_k$ is {\it not} updated,
only the first column of $\signal{U}_k^{(k)}$
(i.e., $\signal{S}_k^T\signal{u}_k$) should be computed.
This is because, when $\signal{S}_k$ is {\it not} updated,
it holds that
$\signal{U}_\iota^{(k)} = \signal{U}_\iota^{(k-1)}$ for
$\iota=\mathcal{I}_k\setminus \{k\}$
and
$[\signal{U}_k^{(k)}]_{2:r} = [\signal{U}_{k-1}^{(k-1)}]_{1:r-1}$,
where $[\signal{A}]_{a:b}$ designates the submatrix of $\signal{A}$
consisting of the $a$th to $b$th column vectors.
Thus, the average complexity for $\signal{U}_\iota^{(k)}$ is
$[(q+r-1)DN + (m-1)DN]/m$.
For the same reason as $(\signal{U}_\iota)_{\iota\in\mathcal{I}_k}$,
the matrices $(\signal{U}_\iota^{(k)})_{\iota\in\mathcal{I}_k}$
also have only $q+r-1$ distinct column vectors,
hence the complexity to compute $\signal{e}_\iota^{(k)}$
and $\signal{a}_\iota^{(k)}$ is no more than $2(q+r-1)D$.
Overall, the total complexity for the filter update is
$\alpha(q,r,m)DN + (4q +2r)D + (r+7)q +2$,
where $\alpha(q,r,m):=(q+r+m-2)/m$.
If we set, for instance, $D=5$, $m=10$, $r=1$, and $q=5$
(which are used in Section \ref{subsec:prop_cg_artificial}),
the complexity for the filter update is $7N+152$.

Finally, we consider the case where $q$ parallel processors are
available.
In this case, the computation of the variables corresponding to
each $\iota\in\mathcal{I}_k$
is naturally assigned to each processor.
We consider the complexity imposed on each processor
at each iteration.
The complexity to compute $\signal{U}_\iota^{(k)}$ is $rDN$,
when $\signal{S}_k$ is updated, and $DN$,
when $\signal{S}_k$ is {\it not} updated.
The average complexity is thus $\beta(r,m)DN$, where $\beta(r,m):=(r+m-1)/m$.
Overall, the per-processor complexity for the filter update is
$\beta(r,m)DN + (2r+4)D +r+9$.
For $D=5$, $m=10$, $r=1$, and an arbitrary $q$,
the complexity for the filter update is $5N + 40$.

\begin{figure}[t]
\centering
 \includegraphics[width=8cm]{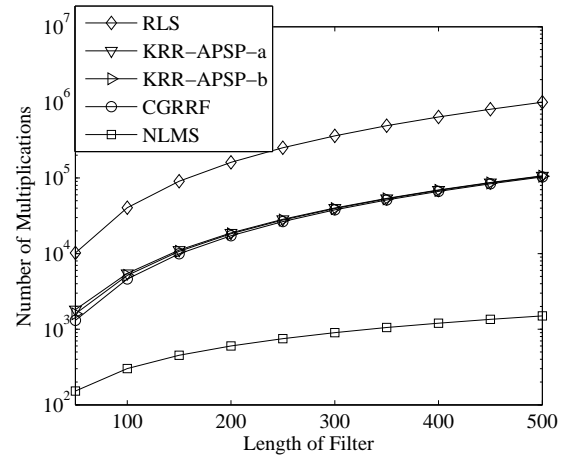}
 \caption{Complexities of the conventional methods and
the proposed algorithm with (a) single
 processor and (b) $q$ processors.}
\label{fig:complexity}
\end{figure}



 \begin{table}[t]
\centering
\caption{Computational complexities of algorithms.}{
\label{table:complexity}
\begin{tabular}{|c|c|}
\hline \rule{0cm}{2.5ex}Algorithm &  Number of multiplications
per iteration  \\ \hline
\emph{\small \bf NLMS} & {\small $3N+2$}                        \\ \hline
\emph{\small \bf RLS} & {\small $4N^2 + 4N + 1$}  \\ \hline
\emph{\small \bf CGRRF} & {\small $(D-1)N^2/m + [(5D-4)/m +4]N $} \\
&  {\small $+ 2(D-1)$} \\ \hline
\emph{\small \bf KRR-APSP} &  {\small $(D-1)N^2/m$} \\
(single processor)&  {\small $+[(5D-4)/m +4]N +\alpha(q,r,m)DN$} \\
&  {\small $+ 2(D-1) + (4q +2r)D + (r+7)q +2$}\\ \hline
\emph{\small \bf KRR-APSP} &  {\small $(D-1)N^2/m$} \\
($q$ processors)&  {\small $+[(5D-4)/m +4]N +\beta(r,m)DN$} \\
&  {\small $+ 2(D-1) + (2r+4)D + r +9$}\\ \hline
\end{tabular}
}
\end{table}

In TABLE \ref{table:complexity}, the overall complexity of
the proposed algorithm is summarized with those of the NLMS algorithm,
the RLS algorithm \cite[Table 9.1]{haykin}, and CGRRF
\cite{chowdhury01}; we assume for fairness that CGRRF updates
the filter every $m$ iterations.
Figure \ref{fig:complexity} plots the number of multiplications
against the filter length $N$ for $D=5$, $m=10$, $r=1$, and $q=5$
(which are used in Section \ref{subsec:prop_cg_artificial}).
We can see that the complexity of the proposed algorithm is much lower than
that of RLS (due to the factor $m$), and marginally higher than that of
CGRRF; in particular, for a large value of $N$, the difference between
the proposed and CGRRF methods is negligible.
Moreover, compared with NLMS, the proposed algorithm requires higher
complexity for realizing better performance.
However, the difference can be significantly reduced by increasing
$m$; in our experiments, the use of $m=100$ gives almost the same
performance as the use of $m=10$.
It should be mentioned that the difference (in computational complexity)
between CGRRF and KRR-APSP can be further reduced by taking
into account the update date of the vector $\widetilde{\signal{h}}_k$
(i.e., the rate in which it happens that
$\norm{\signal{e}_\iota^{(k)}}^2\leq \rho$).
If we choose $\rho$ appropriately, the update rate is typically less
than $10$ \%.

In conclusion, the proposed algorithm is highly expected to realize,
with comparable computational complexity, superior tracking
performance to the existing Krylov-subspace-based reduced-rank methods,
as will be verified by simulations in Section \ref{sec:numer_examp}.
Moreover, the algorithm has a fault tolerance nature
thanks to its inherently parallel structure;
i.e., even if some of the engaged concurrent processors are crashed,
the lack of information from the crashed processors would {\it not}
cause any serious degradation in performance.
This is because the direction of update is determined by taking into
account all the directions suggested by each input data vector little by
little.

In the following section, we present an analysis of the proposed
algorithm.

\section{Analysis of the Proposed Algorithm} \label{sec:analysis}

In the adaptive filtering or learning, the observed measurements are
mostly corrupted by noise and the environments are nonstationary in many
scenarios.
Under such uncertain situations, it is difficult (or nearly impossible)
to guarantee that the adaptive filter approaches the optimal one
monotonically at every iteration.
Thus, a meaningful and realistic property
desired for an adaptive algorithm would be to approach every point in an
appropriately designed set of filtering vectors
monotonically at each iteration.
How can such a set, say $\Omega_k\subset\real^N$, be designed?

In our analysis, we let $\Theta_k:\real^N\rightarrow [0,\infty)$
be a (continuous and convex) objective function, and
$\Omega_k$ is defined as a set of all the vectors that achieve
the infimum of $\Theta_k$ over a certain constraint set.
(The constraint is associated with the requirements that the filter
should lie in the Krylov subspace.)
Then, the desired {\it monotone approximation} property is expressed as
follows\footnote{To ensure
\refeq{eq:mono_appro}, {\it closedness and convexity} of $\Omega_k$ are
essential.}:
\begin{equation}
  \norm{\signal{h}_{k+1} - \signal{h}_{(k)}^*}
\leq \norm{\signal{h}_k -
  \signal{h}_{(k)}^*}, \ \forall \signal{h}_{(k)}^*\in\Omega_k, \
 k\in\Natural.\label{eq:mono_appro}
\end{equation}
We stress that \refeq{eq:mono_appro} insists that the monotonicity
holds {\it for all the elements of} $\Omega_k$.

What about \sinq{optimality} in terms of the objective function
$\Theta_k$?
Is it possible to prove \sinq{optimality} in any sense?
As you might notice, the objective function $\Theta_k$ depends on $k$.
Namely, what we should \sinq{minimize} is {\it not} a fixed objective
function but is a sequence of objective functions
$(\Theta_k)_{k\in\Natural}$.
This is the major difference from the normal optimization problems, and
this formulation naturally fits the adaptive signal processing
because the objective function should be changing in conjunction with
changing environments.
Thus, a meaningful \sinq{optimality} to show would be that
$(\signal{h}_k)_{k\in\Natural}$ minimizes  $(\Theta_k)_{k\in\Natural}$
asymptotically; i.e.,
\begin{equation}
\lim_{k\rightarrow \infty} \Theta_k(\signal{h}_k)=0,\label{eq:asymp_opti}
\end{equation}
which is called {\it asymptotic optimality}
\cite{yamada03_kaisetsu,yagu_paper}.

The goal of this section is to prove that the proposed algorithm enjoys
the two desired properties \refeq{eq:mono_appro} and
\refeq{eq:asymp_opti}.
To this end, we firstly build, with the objective function $\Theta_k$,
a unified framework named {\it reduced-rank adaptive projected
subgradient method (R-APSM)}, and derive the proposed algorithm
from R-APSM with a specific design of $\Theta_k$.
We then prove that R-APSM, including the proposed algorithm as its
special case, has the desired properties under some mild conditions.

 \begin{figure}[t]
\begin{center}
\includegraphics[width=8cm]{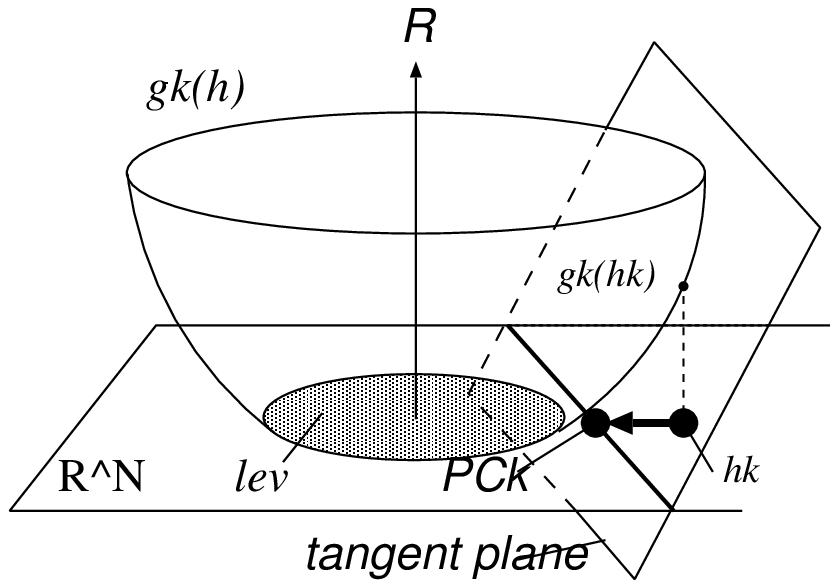}
\end{center}
\caption{A geometric interpretation of
the subgradient projection $T_{{\rm sp}(\Theta_k)}(\signal{h}_k)$
when lev$_{\leq 0}\Theta_k(:=\{\signal{h}\in\real^N:
\Theta_k(\signal{h})\leq 0\})\neq\emptyset$.}
\label{fig:subgradient}
 \end{figure}

\subsection{Alternative Derivation of the Proposed Algorithm}
\label{subsec:derivation}

Recall here that $\signal{h}_k$ is forced to lie in
$\mathcal{R}(\signal{S}_k)$ at each iteration $k\in\Natural$.
For an analysis of the proposed algorithm, we define
\begin{equation}
\signal{\Phi}_k:=\signal{S}_{k+1}\signal{S}_k^T\in \real^{N \times N}.
\end{equation}
Given an arbitrary $\signal{h}_0\in\real^N$ and
a sequence of continuous convex objective functions
$\Theta_k:\real^N\rightarrow [0,\infty)$, $k\in\Natural$,
R-APSM \footnote{The original APSM \cite{yamada03_kaisetsu,yagu_paper}
is obtained by replacing $\signal{\Phi}_k$ in \refeq{eq:vrr_apsm} by a
projection operator onto a closed convex set of an absolute constraint.}
generates a sequence $(\signal{h}_k)_{k\in\Natural}\subset\real^N$ by
\begin{equation}
\signal{h}_{k+1}:=\left\{\begin{array}{ll}
\signal{\Phi}_k\left[
\signal{h}_k-\lambda_k
\displaystyle
\frac{\Theta_k(\signal{h}_k)}{\norm{\Theta'_k(\signal{h}_k)}^2}
\Theta'_k(\signal{h}_k)
\right]
\\
\mbox{if } \Theta'_k(\signal{h}_k) \ne
\signal{0},\\[.5em]
\signal{\Phi}_k\signal{h}_k\\
\mbox{otherwise,}
\end{array} \right.
\label{eq:vrr_apsm}
\end{equation}
where $\lambda_k \in [0,2]$, $k \in\Natural$,
and $\Theta'_k(\signal{h}_k)\in$
$\partial\Theta_k(\signal{h}_k)$ is a {\it subgradient}
of $\Theta_k$ at $\signal{h}_k$ (see Appendix \ref{append:mathdef}).

Suppose that lev$_{\leq 0}\Theta_k:=\{\signal{h}\in\real^N:
\Theta_k(\signal{h})\leq 0\}\neq \emptyset$
($\Leftrightarrow\min_{\signal{h}\in\real^N} \Theta_k(\signal{h})=0$).
Then, removing $\signal{\Phi}_k$,
\refeq{eq:vrr_apsm} for $\lambda_k=1$ is the subgradient projection
relative to $\Theta_k$ [cf.~\refeq{eq:proj_lower}],
which is denoted by $T_{{\rm sp}(\Theta_k)}(\signal{h}_k)$
(see Fig.~\ref{fig:subgradient}).
The update equation in \refeq{eq:vrr_apsm} can be expressed as
 \begin{equation}
  \signal{h}_{k+1} := \signal{\Phi}_k \left[
\signal{h}_k + \lambda_k
\left(
T_{{\rm sp}(\Theta_k)}(\signal{h}_k) - \signal{h}_k
\right)\right].\label{eq:r_apsm2}
 \end{equation}
Noticing that the thick arrow in Fig.~\ref{fig:subgradient}
expresses $T_{{\rm sp}(\Theta_k)}(\signal{h}_k) - \signal{h}_k$,
the figure with \refeq{eq:r_apsm2} provides
a geometric interpretation of R-APSM (except for $\signal{\Phi}_k$).

Let us now derive the proposed algorithm from
R-APSM.
Let $\mathcal{I}_k$ be the control sequence,
and $w_\iota^{(k)}\in(0,1]$,
$\iota\in\mathcal{I}_k$, $k\in\Natural$, the weight, both of which are
defined in the same way as in Section \ref{subsec:strraf}.
An outer approximating closed half-space
$H^-_\iota(\signal{h}_k) \supset C_\iota(\rho)$ is defined as [see \refeq{eq:stochastic}]
\begin{align*}
\hspace*{-.3em} H^-_\iota(\signal{h}_k):= \left\{\signal{h}\in\real^N:
\innerprod{\signal{h} - \signal{h}_k}{\signal{s}_\iota^{(k)}} + g_\iota(\signal{h}_k)\leq 0
\right\},&\\
 \iota\in\mathcal{I}_k, k\in\Natural,&
\end{align*}
where
$\signal{s}_\iota^{(k)}:=\signal{\nabla}g_\iota(\signal{h}_k):=
2\signal{U}_\iota\signal{e}_\iota(\signal{h}_k)
\subset\real^N$.
Because
\begin{enumerate}
 \item $H_\iota^-(\signal{h}_k)$, $\iota\in\mathcal{I}_k$, contains
       favorable vectors because of the definition of $C_\iota(\rho)$,
       and
 \item $\signal{h}_k$ should lie in
    $\mathcal{R}(\signal{S}_k)=\mathcal{K}_D(\widehat{\signal{R}}_k,\widehat{\signal{p}}_k)$,
\end{enumerate}
the distance to $H_\iota^-(\signal{h}_k)\cap
\mathcal{R}(\signal{S}_k)$ is a natural candidate of objective function.
Moreover, for assigning a larger weight to a farther set,
the weight
$d(\signal{h}_k,H^-_\iota(\signal{h}_k)\cap\mathcal{R}(\signal{S}_k))$
is given to the distance function
$d(\signal{h},H^-_\iota(\signal{h}_k)\cap\mathcal{R}(\signal{S}_k))$.
With a normalization factor $L_k:=\sum_{\iota\in\mathcal{I}_k} w_\iota^{(k)}
 d(\signal{h}_k,H^-_\iota(\signal{h}_k)\cap\mathcal{R}(\signal{S}_k))$,
the resulting objective function is given as follows:
\begin{equation}
 \Theta_k(\signal{h}) :=
\left\{
\begin{array}{l}
\displaystyle\frac{1}{L_k}\sum_{\iota\in\mathcal{I}_k} w_\iota^{(k)}
 d(\signal{h}_k,H^-_\iota(\signal{h}_k)\cap\mathcal{R}(\signal{S}_k))\\
\hspace*{5em}\times
 d(\signal{h},H^-_\iota(\signal{h}_k)\cap\mathcal{R}(\signal{S}_k))
 \\
\mbox{if } L_k\neq
 0,\\
0\\
\mbox{otherwise}.
\end{array}
\right.\label{eq:Theta_design}
\end{equation}
An application of R-APSM to $\Theta_k(\signal{h})$
in \refeq{eq:Theta_design} yields (cf.~\cite{yagu_paper})
\begin{eqnarray}
\hspace*{-2em}&&\signal{h}_{k+1} = \nonumber\\
\hspace*{-2em}&&
\signal{\Phi}_k\left[\signal{h}_k + \lambda_k
\mathcal{M}_k \left(\displaystyle\sum_{\iota\in\mathcal{I}_k}
w_\iota^{(k)}
P_{H_\iota^-(\signal{h}_k) \cap
\mathcal{R}(\signal{S}_k)}(\signal{h}_k)
- \signal{h}_k \right)\right],\nonumber\\
\hspace*{-.5em}\label{eq:update_prop_high}
\end{eqnarray}
where $\lambda_k\in [0,2]$, $k\in\Natural$, and
\begin{eqnarray*}
 \mathcal{M}_k:=
\left\{
\begin{array}{l}
1 \quad  \mbox{if }
g_\iota(\signal{h}_k)
 \leq 0, \ \forall \iota\in\mathcal{I}_k, \\
\displaystyle\frac{\displaystyle\sum_{\iota\in\mathcal{I}_k}
w_\iota^{(k)}
\norm{P_{H_\iota^-(\signal{h}_k)\cap
\mathcal{R}(\signal{S}_k)}(\signal{h}_k)
- \signal{h}_k}^2}
{\norm{\displaystyle\sum_{\iota\in\mathcal{I}_k}
w_\iota^{(k)}
P_{H_\iota^-(\signal{h}_k)\cap
\mathcal{R}(\signal{S}_k)}(\signal{h}_k)
- \signal{h}_k}^2}\\
 \mbox{otherwise}.
\end{array}
\right.
\end{eqnarray*}
Noticing $\signal{h}_k\in\mathcal{R}(\signal{S}_k)$ and
defining
$\signal{Q}_k:=\signal{S}_k\signal{S}_k^T$,
the projection of $\signal{h}_k$ onto $H^-_\iota(\signal{h}_k)\cap
 \mathcal{R}(\signal{S}_k)$ is given as follows:
\begin{align}
&P_{H_\iota^-(\signal{h}_k) \cap
 \mathcal{R}(\signal{S}_k)}(\signal{h}_k)
= \nonumber\\
&\left\{
\begin{array}{ll}
\signal{h}_k &  \mbox{if }
g_\iota(\signal{h}_k)
 \leq 0,\\
\signal{h}_k -
 \displaystyle\frac{g_\iota(\signal{h}_k)}
{\norm{\signal{Q}_k\signal{s}_\iota^{(k)}}^2}
\signal{Q}_k\signal{s}_\iota^{(k)}&
 \mbox{otherwise}.
\end{array}
\right.\label{eq:proj_high}
\end{align}
Letting $\signal{h}_k=\signal{S}_k \widetilde{\signal{h}}_k$,
we obtain
$\signal{e}_\iota(\signal{h}_k)=\signal{e}_\iota^{(k)}
(\widetilde{\signal{h}}_k)$,
$g_\iota(\signal{h}_k)=
g_\iota^{(k)}(\widetilde{\signal{h}}_k)$, and
$\signal{S}_k^T\signal{s}_\iota^{(k)} =
\widetilde{\signal{s}}_\iota^{(k)}$,
from which and $P_{H_\iota^-(\signal{h}_k) \cap
 \mathcal{R}(\signal{S}_k)}(\signal{h}_k)\in\mathcal{R}(\signal{S}_k)$
we can verify
\begin{equation}
P_{H_\iota^-(\signal{h}_k) \cap
 \mathcal{R}(\signal{S}_k)}(\signal{h}_k)
=\signal{S}_kP_{\widetilde{H}_{\iota,k}^-(\widetilde{\signal{h}}_k)}(\widetilde{\signal{h}}_k).
\label{eq:high_and_low}
\end{equation}
Substituting \refeq{eq:high_and_low} and $\signal{h}_k=\signal{S}_k \widetilde{\signal{h}}_k$
into \refeq{eq:update_prop_high},
and left-multiplying both sides of \refeq{eq:update_prop_high}
by $\signal{S}_k^T$, we obtain the proposed algorithm.
Taking a look at the update equation in \refeq{eq:update_prop_high},
it is seen that it has the same form as the {\it linearly constrained
adaptive filtering algorithm} \cite{yry_j_ieice05} except for the
mapping $\signal{\Phi}_k$ from $\mathcal{R}(\signal{S}_k)$ to
$\mathcal{R}(\signal{S}_{k+1})$.
Hence, viewing the behavior of the proposed algorithm in
$\real^N$, {\it it performs parallel subgradient projection
in a series of (constraint) Krylov subspaces}
$(\mathcal{R}(\signal{S}_k))_{k\in\Natural}$.


\subsection{Analysis of R-APSM}
\label{subsec:analysis}
We prove that the sequence
$(\signal{h}_k)_{k\in\Natural}$ generated by R-APSM satisfies
the desired properties \refeq{eq:mono_appro} and \refeq{eq:asymp_opti}.
In the analysis, {\it the fixed point set} of the \sinq{mapping}
$\signal{\Phi}_k(:=\signal{S}_{k+1}\signal{S}_k^T):\real^N\rightarrow
\mathcal{R}(\signal{S}_{k+1})$,
$\signal{a}\mapsto \signal{\Phi}_k\signal{a}$, plays an important role.
{\it What is the fixed point set?}
Given a mapping
$T:\real^N\rightarrow\real^N$, a point $\signal{x}\in\real^N$ satisfying $T (\signal{x})=\signal{x}$ is called a {\it
fixed point} of $T$.
Moreover, the set of all such points, i.e. the set
$\Fix{T}:= \left\{\signal{x}\in\real^N: T(\signal{x})=\signal{x}\right\}$, is called
the {\it fixed point set} of $T$.
The set $\Fix{\signal{\Phi}_k}$ is characterized as below.

\begin{proposition}
\label{fact:charac_fixphik}
{\it (Characterizations of $\Fix{\signal{\Phi}_k}$)}
\begin{enumerate}
 \item [(a)] $\signal{0}\in \Fix{\signal{\Phi}_k}$.

 \item [(b)] $\Fix{\signal{\Phi}_k}\subset
       \mathcal{R}(\signal{S}_k) \cap \mathcal{R}(\signal{S}_{k+1})$.

 \item [(c)]
\begin{equation}
\hspace*{-3em}\Fix{\signal{\Phi}_k}
= \left\{ \signal{S}_k \widetilde{\signal{z}} =\signal{S}_{k+1} \widetilde{\signal{z}}:
\widetilde{\signal{z}} \in \Fix{\signal{S}_k^T\signal{S}_{k+1}}\subset \real^D
    \right\}, \label{eq:fact_2c1}
\end{equation}
and
\begin{eqnarray}
 \Fix{\signal{S}_k^T\signal{S}_{k+1}} =
\left\{
\widetilde{\signal{z}}\in\real^D:
 \signal{S}_{k+1} \widetilde{\signal{z}} = \signal{S}_k \widetilde{\signal{z}}
\right\}.\label{eq:fixsksk1}
\end{eqnarray}

 \item [(d)] If $\signal{S}_{k+1}= \signal{S}_k$, then
$\signal{\Phi}_k = P_{\mathcal{R}(\signal{S}_k)}$ and
       $\Fix{\signal{\Phi}_k}=\mathcal{R}(\signal{S}_k)$.

\end{enumerate}
 \end{proposition}
{\it Proof:} See Appendix \ref{append:fact_fixphik}.\migip

Define
\begin{align}
\Theta_k^*:=&\displaystyle\inf_{\signal{x}\in \Fix{\signal{\Phi}_k}}
\Theta_k(\signal{x}), \ k\in\Natural,\\
 \Omega_k:=&\big\{
\signal{h}\in \Fix{\signal{\Phi}_k}:\Theta_k(\signal{h}) =
\Theta_k^*
\big\}, \ k\in\Natural.
\end{align}
(As mentioned before \refeq{eq:mono_appro}, the constraint set
$\Fix{\signal{\Phi}_k}$ is associated with the requirements
$\signal{h}_k\in\mathcal{R}(\signal{S}_k)$ for any $k\in\Natural$.)
Then, the following theorem holds.
\begin{theorem}
\label{prop:optimality}
The sequence $\left(\signal{h}_k\right)_{k\in\Natural}$ generated
 by R-APSM satisfies the following.
\begin{enumerate}
 \item [(a)] (Monotone Approximation)
\begin{enumerate}
 \item [(I)]
Assume $\Omega_k\neq\emptyset$.
Then, for any $\lambda_k\in
       \Big[0,2\left(1-\Theta_k^*/\Theta_k(\signal{h}_k)\right)\Big]$,
\refeq{eq:mono_appro} holds.
 \item [(II)]
Assume in addition
$\Theta_k(\signal{h}_k)>\inf_{\signal{x}\in\real^N}
       \Theta_k(\signal{x})\geq 0$.
Then, for any
       $\lambda_k\in
       \Big(0,2\left(1-\Theta_k^*/\Theta_k(\signal{h}_k)\right)\Big)$,
\begin{equation}
 \norm{\signal{h}_{k+1} - \signal{h}_{(k)}^*} < \norm{\signal{h}_k -
  \signal{h}_{(k)}^*}, \ \forall \signal{h}_{(k)}^*\in\Omega_k.
\label{eq:monotone_aII}
\end{equation}

\end{enumerate}

 \item [(b)] (Boundedness, Asymptotic Optimality)
Assume
\begin{equation}
\exists K_0\in\Natural \mbox{ s.t.~}
\left\{
\begin{array}{l}
\mbox{(i) }
\Theta_k^*=0, \ \forall k\geq K_0, \mbox{ and}\\
\mbox{(ii) }
\Omega:=\bigcap_{k\geq
       K_0}\Omega_k\neq \emptyset.
\end{array}
\right.\label{eq:mono_app_assump_aI}
\end{equation}
Then $(\signal{h}_k)_{k\in\Natural}$ is bounded.
In particular, if there exist $\varepsilon_1,\varepsilon_2>0$ such that
$\lambda_k\in[\varepsilon_1,2 - \varepsilon_2]\subset(0,2)$,
then \refeq{eq:asymp_opti} holds,
provided that  $\left(\Theta'_k(\signal{h}_k)\right)_{k\in\Natural}$ is bounded.
\end{enumerate}
 \end{theorem}
{\it Proof:} See Appendix \ref{append:prop_optimality}.\migip

Finally, for the $\Theta_k$ specified by \refeq{eq:Theta_design},
we discuss the assumptions made in Theorem \ref{prop:optimality}.
First, it is worth mentioning that $\signal{S}_k$ tends to stop moving
when the estimates of $\signal{R}$ and $\signal{p}$ become reliable,
and, in such a case, Proposition \ref{fact:charac_fixphik} implies
$\Fix{\signal{\Phi}_k}=\mathcal{R}(\signal{S}_k)$.
Hence, we assume
$\Fix{\signal{\Phi}_k}=\mathcal{R}(\signal{S}_k)$ for
simplicity here.
Moreover, it mostly holds that
$\bigcap_{\iota\in\mathcal{I}_k}H_\iota^-(\signal{h}_k) \cap
\mathcal{R}(\signal{S}_k)\neq \emptyset$ at each $k\in\Natural$,
unless the observed data are highly inconsistent.
In this case, ($\Theta_k^*=0$ and)
$\Omega_k=\bigcap_{\iota\in\mathcal{I}_k}H_\iota^-(\signal{h}_k) \cap
\mathcal{R}(\signal{S}_k)(\neq \emptyset)$, thus
\refeq{eq:mono_appro} holds.
We remark that, under
$\Fix{\signal{\Phi}_k}=\mathcal{R}(\signal{S}_k)$,
the condition $\bigcap_{\iota\in\mathcal{I}_k}H_\iota^-(\signal{h}_k) \cap
\mathcal{R}(\signal{S}_k)\neq \emptyset$ is sufficient but not
necessary for \refeq{eq:mono_appro} to hold.
(In fact, $\Omega_k$ can be nonempty even if
$\bigcap_{\iota\in\mathcal{I}_k}H_\iota^-(\signal{h}_k) = \emptyset$.)

Under $\Fix{\signal{\Phi}_k}=\mathcal{R}(\signal{S}_k)$,
the conditions in \refeq{eq:mono_app_assump_aI} are satisfied
when
$\bigcap_{k\geq K_0}
\left[
\bigcap_{\iota\in\mathcal{I}_k}H_\iota^-(\signal{h}_k)
\cap \mathcal{R}(\signal{S}_k)
\right] \neq \emptyset$, which mostly holds
if the observed data are consistent for $k\geq K_0$.
We mention that $\left(\Theta'_k(\signal{h}_k)\right)_{k\in\Natural}$
for the $\Theta_k$ in \refeq{eq:Theta_design} is automatically bounded
\cite{yamada_asilomar03}.

In dynamic environments, it is hardly possible to ensure
$\Fix{\signal{\Phi}_k}=\mathcal{R}(\signal{S}_k)$ for all
$k\geq K_0$,
since $\signal{S}_k$ will move when the environments change.
In this case, the asymptotic optimality is difficult to be guaranteed.
However, it is possible that the monotone approximation is guaranteed,
because the environments would be nearly static in some (short)
periods and, within such periods, $\signal{S}_k$ may stop moving.


\section{Numerical Examples}
\label{sec:numer_examp}


\begin{figure}[t!]
\centering
\subfigure[]{
 \includegraphics[width=7cm]{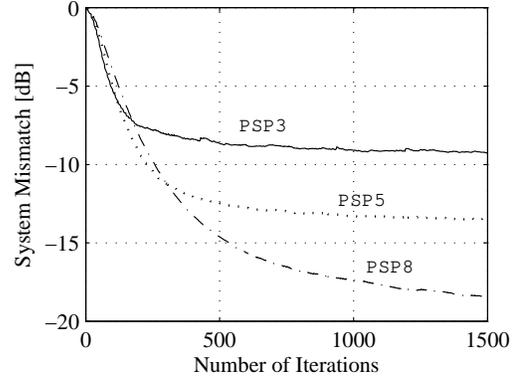}
}
\subfigure[]{
 \includegraphics[width=7cm]{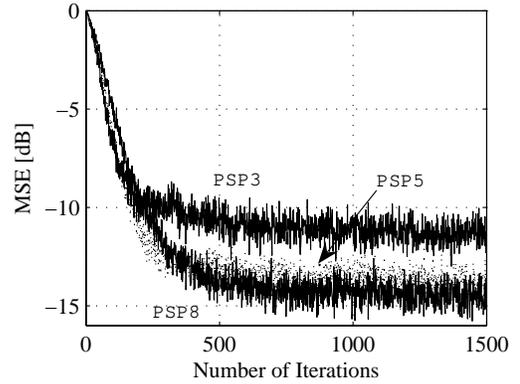}
}
 \caption{Performance of the proposed algorithm for
 $D=3,5,8$, $q=4$, and $r=1$ under SNR $=15$ dB
in (a) system mismatch and (b) MSE.}
\label{fig:psp}
\end{figure}

This section provides numerical examples to verify the advantages of the
proposed algorithm over the CGRRF method
\cite{chowdhury01} (Note: we omit a comparison with the RLS algorithm,
because it is known that CGRRF provides convergence comparable to
RLS with lower computational complexity
and it does not suffer from any numerical instability problems
\cite{boray92,chang00}).
In the current study,
weakly correlated input signals are employed in order to avoid
preconditioning for conciseness.
In simple system identification problems,
we firstly examine the performance of the proposed algorithm for
different values of $D$ and $q$, and then compare the proposed algorithm
with CGRRF.
We finally apply the two methods to a multiple access interference
suppression problem in code-division multiple-access
(CDMA) wireless communication systems.
In all the simulations, we set $\mathcal{I}_k:=\{k,k-1,\cdots,k-q+1\}$,
and the matrix $\signal{S}_k$ is updated
every $m=10$ iterations
with $\widehat{\signal{R}}_0:=\signal{O}$,
$\widehat{\signal{p}}_0:=\signal{0}$, and $\gamma=0.999$.

\subsection{Performance of the Proposed Algorithm for
System Identification}\label{subsec:Dq}

\begin{figure}[t!]
\centering
\subfigure[]{
 \includegraphics[width=7cm]{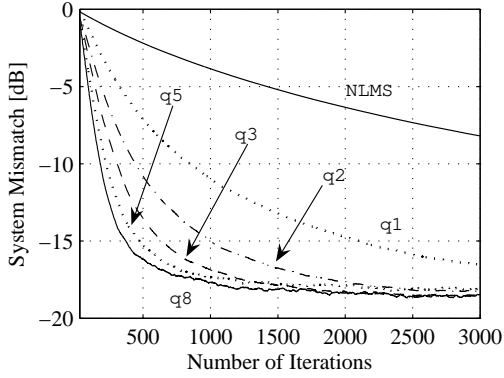}
}
\subfigure[]{
 \includegraphics[width=7cm]{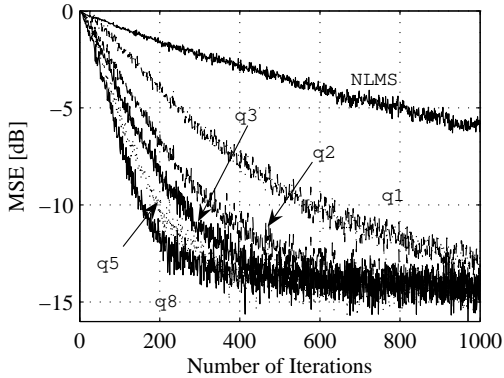}
}
 \caption{Performance of the proposed algorithm for
 $D=8$, $q=1,2,3,5,8$, and $r=1$ under SNR $=15$ dB in (a) system mismatch and (b)
 MSE.}\vspace{-1em}
\label{fig:dif_q}
\end{figure}

To compute arithmetic averages of MSE and system mismatch,
i.e.~$\norm{\signal{h}^* - \signal{h}_k}^2/\norm{\signal{h}^*}^2$,
$300$ independent experiments are performed.
In each experiment, $\signal{h}^*$ is generated randomly for $N=50$, and
the input signal is generated by passing a white Gaussian signal through
a length-$30$ finite impulse response (FIR) filter whose coefficients
are chosen randomly (the resulting input signal has weak autocorrelation).
The signal to noise ratio (SNR) is set to
SNR $:=10\log_{10}\left(E\left\{z_k^2\right\}/E
\left\{n_k^2\right\}\right)=15$ dB,
where $z_k:=\innerprod{\signal{u}_k}{\signal{h}^*}$.

The parameters are set to\footnote{
In the current study, we only focus on the case of $r=1$
to make the parameter settings simple.
In fact, it has been reported in
\cite{yry_j_ieice05,yuya_power_ieee,yumuya_eurasip,ysy_apqp2007,cavalcante08}
that fast convergence and good steady-state performance are
attained when we use $r=1$ and a large value of $q$ (e.g., $q=8,16,32$)
for the $N$ within the range of $64$ to $2000$
in the (full-rank) APSP algorithm \cite{ysy.sp}.
}
$\lambda_k=0.03$,
$\rho=0.15$, $q=4$, $r=1$, $\widetilde{\signal{h}}_0=\signal{0}$, and
$D=3,5,8$.
The results are depicted in Fig.~\ref{fig:psp}.
It is seen that, from $D=3$ to $D=5$, an increase of $D$ leads to better
steady-state performance both in system mismatch and MSE.
However, from $D=5$ to $D=8$, the gain in MSE is slight, although
a significant gain is obtained in system mismatch.
This is because the value of $\norm{\signal{h}_k - \signal{h}^*}$ at the
steady state is still not small enough in the case of $D=5$, but
the value of $\norm{\signal{h}_k - \signal{h}^*}_{\signal{R}}$ is
already small enough (see Section \ref{sec:motivation}).

Next we fix the value of $D=8$, and change the value of $q$ as
$q=1,2,3,5,8$.
The rest of the parameters are the same as in Fig.~\ref{fig:psp}.
The results are depicted in Fig.~\ref{fig:dif_q}.
As a benchmark, the performance curves of NLMS for step size
$\lambda_k=0.03$ are also drawn.
It is seen that an increase of $q$ (the number of parallel projections
computed at each iteration) raises the speed of convergence
significantly.

\begin{figure}[t!]
\centering
\subfigure[]{
 \includegraphics[width=8cm]{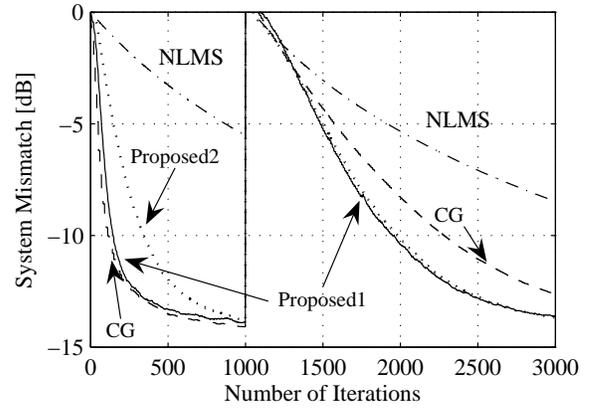}
}
\subfigure[]{
 \includegraphics[width=8cm]{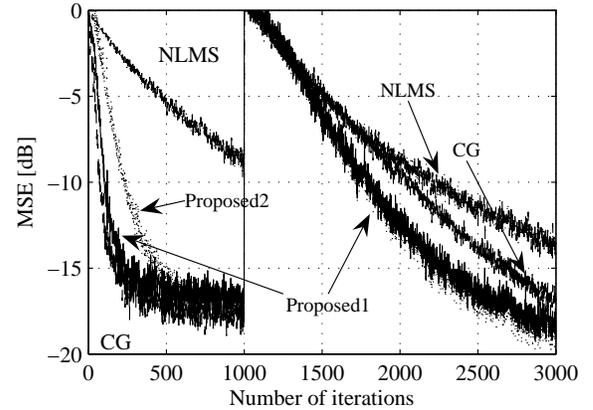}
}
 \caption{The proposed algorithm versus CGRRF and  NLMS under SNR $=20$
 dB in (a) system mismatch and (b) MSE.
For the proposed algorithm, $\lambda_k=0.05$, $k\in\Natural$, $D=5$,
 $\rho=0.1$, and $r=1$.
For CGRRF, $D=5$. For NLMS, $\lambda_k=0.05$, $k\in\Natural$.
}\vspace{-1em}
\label{fig:err_hoptchange}
\end{figure}

\subsection{Proposed versus CGRRF for System Identification}
\label{subsec:prop_cg_artificial}

We compare the performance of the proposed algorithm with CGRRF
and the NLMS algorithm.
The $\signal{h}^*$ and the input signals are generated in the same way
as in Section \ref{subsec:Dq}, and the SNR is set to SNR $=20$ dB.
We consider the situation where $\signal{h}^*$ changes dynamically at
$1000$th iteration; the input statistics are {\it unchanged}, which
means that only the crosscorrelation vector $\signal{p}$ is changed.
For all the algorithms (except for CGRRF), the step size is set
to  $\lambda_k=0.05$, and for the proposed algorithm, we set
$\rho=0.1$, $q=1,5$, $r=1$, $\widetilde{\signal{h}}_0=\signal{0}$, and
$D=5$.
For CGRRF, the Krylov subspace dimension is set also to $D=5$,
and the initial vector at each time instant is set to the zero vector.

Figure \ref{fig:err_hoptchange} plots the results.
As expected from the discussion in Section
\ref{sec:motivation}, the tracking speed of CGRRF after the
sudden change of $\signal{h}^*$ is slow, although its convergence speed
at the initial phase is fast.
On the other hand, the proposed algorithm for $q=5$ achieves fast
initial convergence and good tracking performance simultaneously.

\begin{figure}[t!]
\centering
 \includegraphics[width=9cm]{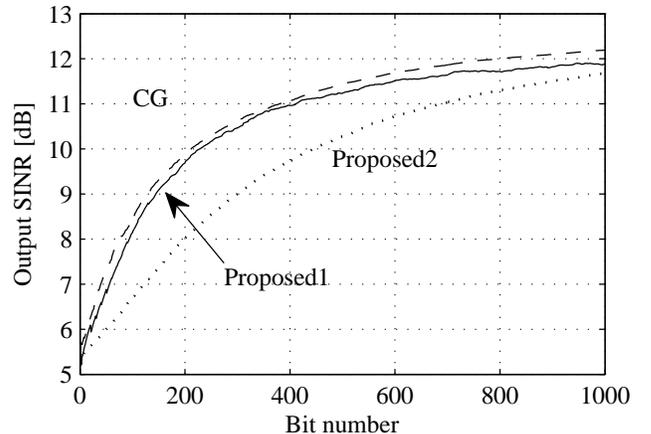}
 \caption{Interference suppression capability in CDMA systems under
SNR $=15$ dB in static environments.
The number of users is $K=8$, and the amplitudes of all users are
 equal.
For the proposed algorithm, $\lambda_k=0.02$, $k\in\Natural$, $D=5$,
 $\rho=0.01$, and $r=1$.
For CGRRF, $D=5$.
}\vspace{-1em}
\label{fig:cdma}
\end{figure}

\begin{figure}[t!]
\centering 
 \includegraphics[width=9cm]{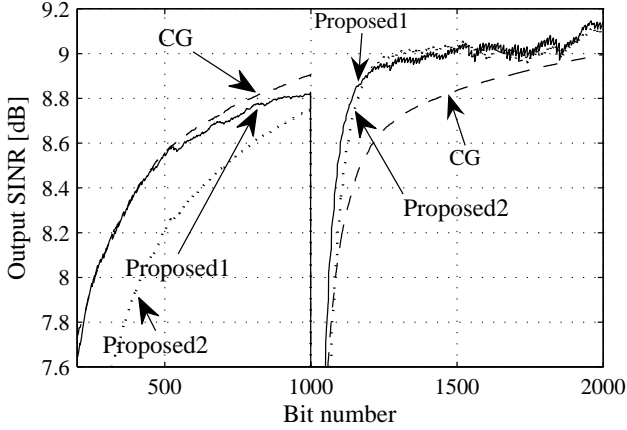}
 \caption{Interference suppression capability in CDMA systems under
SNR $=10$ dB in dynamic environments.
The number of users is changed at the bit number $1000$ from $K=4$ to
 $K=2$.
For the proposed algorithm, $\lambda_k=0.02$, $k\in\Natural$, $D=5$,
 $\rho=0.1$, and $r=1$.
For CGRRF, $D=5$.
}\vspace{-1em}
\label{fig:cdma_rhochange}
\end{figure}

\subsection{Proposed versus CGRRF for Interference Suppression Problem
in CDMA Systems}
\label{subsec:cdma_simulations}
We apply the proposed algorithm and CGRRF to the multiple access
interference suppression problem occurring in the CDMA systems
(see, e.g., \cite{madhow94}).
The received data vector, corresponding to the input vector
$\signal{u}_k$, is given as
\begin{equation}
 \signal{u}_k := \signal{S}\signal{A} \signal{b}_k + \signal{w}_k.
\end{equation}
Here, letting $K$ denote the number of users accessing the same channel,
$\signal{S}\in\real^{N\times K}$ is the signature matrix (each column
corresponds to each user),
$\signal{A}\in\real^{K\times K}$ a diagonal matrix with the amplitudes
from the $K$ users, $\signal{b}_k\in\{1,-1\}^K$ the data symbol vector
of the $K$ users, and $\signal{w}_k\in\real^N$ the vector of additive
white Gaussian noise with zero mean.
The output $d_k$ in Fig.~\ref{fig:rrfilter} corresponds to
the element of $\signal{b}_k$ associated with the desired user.
For simplicity, we assume chip-synchronous but code-asynchronous
systems, as usual in the literature on this problem,
and fading of the channels is not considered.
Also we assume that the training sequence is available to adapt the
filter $\signal{h}_k$.
For the spreading codes, the length-$31$ Gold sequences are employed
(i.e., $N=31$).

In the first simulation, we assume static environments
with $K=8$ users having equal amplitudes under SNR $=15$ dB.
We set $D=5$ for both CGRRF and the proposed algorithm,
and $\lambda_k=0.02$, $\rho=0.01$, $r=1$, and $q=1,5$
for the proposed algorithm.
At the iteration $k=0$, the rank-reduction matrix
$\signal{S}_1\in\real^{N\times D}$ is firstly computed, and then
the lower-dimensional adaptive filter $\widetilde{\signal{h}}_k$ is
initialized as $\widetilde{\signal{h}}_0:= \signal{S}_1^T\signal{s}$,
where $\signal{s}\in\real^N$ is the signature vector of the desired
user.
For CGRRF, the initial vector at each time instant is set to
$\signal{s}$.
The results are depicted in Fig.~\ref{fig:cdma}.

In the second simulation, we assume dynamic environments under
SNR $=10$ dB.
At the beginning, there are $K=4$ users accessing the same channel
simultaneously, and, at the bit number $1000$, all the interfering users
stop their access and another interfering user establishes a new
connection to the channel (i.e., the total number of accessing users
after the bit number $1000$ is $K=2$).
All the interfering signals have twice larger amplitudes than
the desired one.
For the proposed algorithm, we set $\rho=0.1$ and the other parameters
are the same as in the first simulation.
The parameters for CGRRF are the same as in the first
simulation.
The results are depicted in Fig.~\ref{fig:cdma_rhochange}.

From Fig.~\ref{fig:cdma}, it is seen that the proposed algorithm
(for $q=5$) performs similarly to CGRRF in the static environments.
From Fig.~\ref{fig:cdma_rhochange}, on the other hand,
it is seen that the proposed algorithm exhibits better tracking
performance than CGRRF.
This is consistent with the results in Fig.~\ref{fig:err_hoptchange}
and also with the discussion in Section \ref{sec:motivation}.

\section{Conclusion}\label{sec:conclusion}
This paper has presented a robust reduced-rank adaptive filtering
algorithm based on the Krylov subspace and the set-theoretic
adaptive filtering method.
The proposed algorithm provides excellent tradeoff between
performance (in particular, tracking capability) and computational
complexity.
The valuable properties (monotone approximation and asymptotic optimality)
of the proposed algorithm have been proven
within the framework of the modified APSM.
It would be worth repeating that the algorithm has
a fault tolerance nature due to its inherently parallel structure.
The numerical examples have demonstrated that the proposed algorithm
exhibits much better tracking performance than CGRRF
(with comparable computational complexity).
This suggests that the proposed algorithm should perform
better than the existing Krylov-subspace-based reduced-rank methods
in nonstationary environments.
We finally mention that the proposed algorithm
has no numerical problems, since it requires no matrix
inversion, which implies that the algorithm is easy to implement.

\appendices

\newcounter{appnum}
\setcounter{appnum}{1}

\setcounter{proposition}{0}
\renewcommand{\theproposition}{\Alph{appnum}.\arabic{proposition}}
\setcounter{lemma}{0}
\renewcommand{\thelemma}{\Alph{appnum}.\arabic{lemma}}
\setcounter{fact}{0}
\renewcommand{\thefact}{\Alph{appnum}.\arabic{fact}}
\setcounter{example}{0}
\renewcommand{\theexample}{\Alph{appnum}.\arabic{example}}
\setcounter{equation}{0}
\renewcommand{\theequation}{\Alph{appnum}.\arabic{equation}}
\setcounter{claim}{0}
\renewcommand{\theclaim}{\Alph{appnum}.\arabic{claim}}
\setcounter{remark}{0}
\renewcommand{\theremark}{\Alph{appnum}.\arabic{remark}}

\setcounter{appnum}{1}

\setcounter{equation}{0}
\renewcommand{\theequation}{\Alph{appnum}.\arabic{equation}}

\section{Mathematical Definitions} \label{append:mathdef}

Let $\euclidspace$ denote a real Hilbert space equipped with an inner
product $\innerprod{\cdot}{\cdot}$ and its induced norm $\norm{\cdot}$.
We introduce some mathematical definitions used in this paper.

\begin{enumerate}
 \item[(a)]
A set  $C\subset\euclidspace$ is said
to be {\it convex} if $\nu\signal{x}+(1-\nu)\signal{y} \in
C$, $\forall\signal{x},\signal{y}\in C$,
$\forall \nu \in (0,1)$.
A function $\Theta:\euclidspace\rightarrow\real$ is said to be {\it
convex} if $\Theta(\nu\signal{x}+(1-\nu)\signal{y}) \le \nu
\Theta(\signal{x})+(1-\nu)\Theta(\signal{y})$,
 $\forall\signal{x},\signal{y}\in \euclidspace$,
$\forall \nu \in (0,1)$; the inequality is sometimes called {\it
       Jensen's inequality} \cite{boyd04_convexbook}.

 \item[(b)]
A mapping $T$ is said to be
(i) {\it nonexpansive} if $\norm{T(\signal{x}) - T(\signal{y})}\leq
 \norm{\signal{x} - \signal{y}}$, $\forall
 \signal{x},\signal{y}\in\euclidspace$;
(ii) {\it attracting nonexpansive} if $T$ is nonexpansive with $\Fix{T}\neq
\emptyset$ and
$\norm{T(\signal{x}) - \signal{f}}^2 < \norm{\signal{x} -
\signal{f}}^2$,
$\forall (\signal{x},\signal{f})\in\euclidspace\setminus\Fix{T}
\times \Fix{T}$; and
(iii) {\it strongly} or  $\eta$-{\it attracting nonexpansive}
if $T$ is nonexpansive with $\Fix{T}\neq \emptyset$ and
there exists $\eta>0$  s.t.~$\eta\norm{\signal{x} - T(\signal{x})}^2
\leq \norm{\signal{x} - \signal{f}}^2 - \norm{T(\signal{x}) -
\signal{f}}^2$, $\forall \signal{x}\in\euclidspace$, $\forall \signal{f}
\in \Fix{T}$.

 \item[(c)]
Given a continuous convex function
 $\Theta:\euclidspace\rightarrow\real$, the {\it subdifferential} of $\Theta$
at any $\signal{\signal{y}}\in\euclidspace$, defined as
$\partial \Theta(\signal{y}):=\{\signal{a} \in \euclidspace :
\innerprod{\signal{x}-\signal{y}}{\signal{a}}
+ \Theta(\signal{y})
\le \Theta(\signal{x}), \forall
\signal{x} \in \euclidspace \}$, is nonempty.
An element of the subdifferential $\partial \Theta(\signal{y})$
is called a subgradient of $\Theta$ at $\signal{y}$.

 \item[(d)]
Suppose that a continuous convex function $\Theta:\euclidspace\rightarrow \real$
satisfies lev$_{\leq 0} \Theta:=\{\signal{x}\in\euclidspace:
\Theta(\signal{x})\leq 0\}\neq \emptyset$.
Then, for a subgradient $\Theta'(\signal{x})\in \partial\Theta(\signal{x})$,
a mapping $T_{\rm sp(\Theta)}:\euclidspace \rightarrow \euclidspace$
defined by
\begin{eqnarray*}
 T_{\rm sp(\Theta)}(\signal{x}) :=
\left\{
\begin{array}{ll}
\signal{x} -
\displaystyle \frac{\Theta(\signal{x})}{\norm{\Theta'(\signal{x})}^2}\Theta'(\signal{x})
& \mbox{if } \Theta(\signal{x})>0\\
\signal{x} & \mbox{if } \Theta(\signal{x})\leq 0\\
\end{array}
\right.
\end{eqnarray*}
is called {\it a subgradient projection relative to} $\Theta$
(see, e.g., \cite{yagu_paper}).

\end{enumerate}

 \setcounter{appnum}{2}
 \setcounter{equation}{0}
 \setcounter{fact}{0}

\section{Properties of $\signal{\Phi}_k$ and
Proof of Proposition \ref{fact:charac_fixphik}}
\label{append:fact_fixphik}

This appendix presents basic properties of $\signal{\Phi}_k$,
the proof of Proposition \ref{fact:charac_fixphik}, and some results regarding
the attracting nonexpansivity of $\signal{\Phi}_k$ (see Appendix
\ref{append:mathdef}).
\begin{lemma}
\label{fact:properties}
{\it (Basic properties of $\signal{\Phi}_k$)}
 \begin{enumerate}
  \item [(a)] $\signal{\Phi}_k \signal{x} = \signal{S}_{k+1}
    \widetilde{\signal{x}}$ for all
$\widetilde{\signal{x}}\in\real^D$ and
$\signal{x}=\signal{S}_k\widetilde{\signal{x}}$.

  \item [(b)] For any
    $\signal{x}\in\real^N$,
$\norm{\signal{\Phi}_k\signal{x}} \leq \norm{\signal{x}}$; the equality
    holds if and only if
    $\signal{x}\in\mathcal{R}(\signal{S}_k)$.
    Moreover, the mapping $\signal{\Phi}_k$ is {\it nonexpansive}
    (cf.~Appendix \ref{append:mathdef}).
\migip
 \end{enumerate}
\end{lemma}

\noindent{\it Proof of Lemma \ref{fact:properties}.a}: For all
$\widetilde{\signal{x}}\in\real^D$, we have
$\signal{\Phi}_k\signal{x}=\signal{S}_{k+1}\signal{S}_k^T\signal{S}_k\widetilde{\signal{x}}
= \signal{S}_{k+1}\widetilde{\signal{x}}$.

\noindent{\it Proof of Lemma \ref{fact:properties}.b}:
 $\signal{S}_{k+1}^T\signal{S}_{k+1}=\signal{S}_k^T\signal{S}_k=\signal{I}$,
we have, for any $\signal{x}\in\real^N$,
\begin{eqnarray}
 \norm{\signal{\Phi}_k\signal{x}}
&= &
 \norm{\signal{S}_{k+1}\signal{S}_k^T\signal{x}} \nonumber\\
&= &
 \norm{\signal{S}_{k}\signal{S}_k^T\signal{x}}\nonumber\\
&= &
 \norm{P_{\mathcal{R}(\signal{S}_k)} (\signal{x})}\nonumber\\
&\leq &  \norm{\signal{x}}.\label{eq:norm_less}
\end{eqnarray}
The inequality is verified by the nonexpansivity of the projection
operator;
the equality holds if and only if
$\signal{x}\in \mathcal{R}(\signal{S}_k)$.
\refeq{eq:norm_less} and the linearity of $\signal{\Phi}_k$
suggest the nonexpansivity of $\signal{\Phi}_k$.
\migip

\noindent {\bf \underline{Proof of Proposition \ref{fact:charac_fixphik}}}\\
\noindent{\it Proof of Proposition \ref{fact:charac_fixphik}.a}: $\signal{\Phi}_k\signal{0}=\signal{0}$
implies $\signal{0}\in \Fix{\signal{\Phi}_k}$.

\noindent{\it Proof of Proposition \ref{fact:charac_fixphik}.b}: Suppose $\signal{h}\in\Fix{\signal{\Phi}_k}$.
Then, $\signal{h} = \signal{\Phi}_k\signal{h}\in\mathcal{R}(\signal{S}_{k+1})$.
Moreover, by Lemma \ref{fact:properties}.b, $\signal{\Phi}_k\signal{h}
=\signal{h}\Rightarrow
\signal{h}\in\mathcal{R}(\signal{S}_k)$.
Hence $\signal{h}\in\mathcal{R}(\signal{S}_k)\cap
\mathcal{R}(\signal{S}_{k+1})$, implying that
$\Fix{\signal{\Phi}_k}\subset\mathcal{R}(\signal{S}_k)\cap
\mathcal{R}(\signal{S}_{k+1})$.

\noindent{\it Proof of Proposition \ref{fact:charac_fixphik}.c}:
To prove \refeq{eq:fixsksk1}, it is sufficient to show
\begin{equation}
 \signal{S}_k^T\signal{S}_{k+1}\widetilde{\signal{z}} =\widetilde{\signal{z}}
\Leftrightarrow \signal{S}_{k+1}\widetilde{\signal{z}} = \signal{S}_k\widetilde{\signal{z}}.\label{eq:goal21}
\end{equation}
Assume  $\signal{S}_k^T\signal{S}_{k+1}\widetilde{\signal{z}} =\widetilde{\signal{z}}$.
Then, we have
\begin{eqnarray}
\hspace*{-2em}&&\signal{S}_k \signal{S}_k^T\signal{S}_{k+1}\widetilde{\signal{z}}
 =\signal{S}_k\widetilde{\signal{z}} \nonumber\\
\hspace*{-2em}&\Leftrightarrow& P_{\mathcal{R}(\signal{S}_k)}
 \left(\signal{S}_{k+1}\widetilde{\signal{z}}\right)
=\signal{S}_k\widetilde{\signal{z}} \label{eq:proof_c1}\\
\hspace*{-2em}&\Rightarrow& \norm{P_{\mathcal{R}(\signal{S}_k)}
 \left(\signal{S}_{k+1}\widetilde{\signal{z}}\right)} =
 \norm{\signal{S}_k\widetilde{\signal{z}}} =
\norm{\widetilde{\signal{z}}} =
 \norm{\signal{S}_{k+1}\widetilde{\signal{z}}}\label{eq:proof_tmp1}\\
\hspace*{-2em}&\Leftrightarrow& \norm{P_{\mathcal{R}(\signal{S}_k)}
 \left(\signal{S}_{k+1}\widetilde{\signal{z}}\right) -
 \signal{S}_{k+1}\widetilde{\signal{z}}}=0
\label{eq:proof_tmp2}\\
\hspace*{-2em}&\Leftrightarrow& P_{\mathcal{R}(\signal{S}_k)}
 \left(\signal{S}_{k+1}\widetilde{\signal{z}}\right)
=\signal{S}_{k+1}\widetilde{\signal{z}}.\label{eq:proof_c2}
\end{eqnarray}
Here, the equivalence between \refeq{eq:proof_tmp1} and
\refeq{eq:proof_tmp2} is verified
by the well-known Pythagorean theorem.
From \refeq{eq:proof_c1} and  \refeq{eq:proof_c2}, we obtain
$\signal{S}_{k+1}\widetilde{\signal{z}} = \signal{S}_k\widetilde{\signal{z}}$.
The converse is obvious, which verifies \refeq{eq:goal21}.

By Proposition \ref{fact:charac_fixphik}.b, any element
$\signal{z}\in\Fix{\signal{\Phi}_k}$
can be expressed as $\signal{z}=\signal{S}_{k+1}\widetilde{\signal{z}}$,
$\exists \widetilde{\signal{z}}\in\real^D$.
Then, we have
\begin{eqnarray}
\signal{S}_{k+1}\widetilde{\signal{z}} \in\Fix{\signal{\Phi}_k}
&\Leftrightarrow& \signal{S}_{k+1}
 \signal{S}_k^T\signal{S}_{k+1}\widetilde{\signal{z}} =\signal{S}_{k+1}\widetilde{\signal{z}}
\nonumber\\
&\Leftrightarrow& \signal{S}_k^T\signal{S}_{k+1}\widetilde{\signal{z}} =\widetilde{\signal{z}}\nonumber\\
&\Leftrightarrow& \widetilde{\signal{z}}\in\Fix{\signal{S}_k^T\signal{S}_{k+1}},
 \label{eq:sec_equiv}
\end{eqnarray}
which with \refeq{eq:fixsksk1} verifies \refeq{eq:fact_2c1}.

\noindent{\it Proof of Proposition \ref{fact:charac_fixphik}.d}:
The orthonormality of $\signal{S}_k$ and $\signal{S}_k=\signal{S}_{k+1}$
imply that
$\signal{\Phi}_k=P_{{\mathcal{R}(\signal{S}_k)}}$
\cite{strang88}.
Moreover, due to the basic property of projection, we obtain
$\Fix{\signal{\Phi}_k}=
\Fix{P_{{\mathcal{R}(\signal{S}_k)}}}=
\mathcal{R}(\signal{S}_k)$.
\migip


Finally, thanks to Proposition \ref{fact:charac_fixphik}, we can show
that $\signal{\Phi}_k$ is attracting nonexpansive
if and only if $\signal{S}_k=\signal{S}_{k+1}$, as described below.
\begin{lemma}[On attracting nonexpansivity of $\signal{\Phi}_k$]
\label{fact:attracting_ne}
~
\begin{enumerate}
 \item[(a)] If $\signal{S}_k=\signal{S}_{k+1}$, then
 $\signal{\Phi}_k$ is the projection matrix thus {\it 1-attracting
        nonexpansive}.
 \item[(b)] If $\signal{S}_k\neq\signal{S}_{k+1}$, then
 $\signal{\Phi}_k$ is nonexpansive but {\it not} attracting nonexpansive.
\end{enumerate}
\end{lemma}

\noindent{\it Proof of Lemma \ref{fact:attracting_ne}.a}:
By Proposition \ref{fact:charac_fixphik}.d,
$\signal{S}_k=\signal{S}_{k+1}\Rightarrow\signal{\Phi}_k=P_{\mathcal{R}(\signal{S}_k)}$,
$\mathcal{R}(\signal{S}_k)=\Fix{\signal{\Phi}_k}$.
Hence, by the Pythagorean theorem, we have
\begin{align}
\norm{\signal{x} - \signal{\Phi}_k\signal{x}}^2 =
\norm{\signal{x} - \signal{f}}^2 -
\norm{\signal{\Phi}_k\signal{x} -\signal{f}}^2,&\nonumber\\
\forall \signal{x}\in\real^N, \
\forall \signal{f}\in\Fix{\signal{\Phi}_k}.&
\end{align}
This means that the mapping $\signal{\Phi}_k$ is
$1$-attracting nonexpansive.

\noindent{\it Proof of Lemma \ref{fact:attracting_ne}.b}:
By  $\signal{S}_k\neq\signal{S}_{k+1}$,
there exists $\widetilde{\signal{z}}^*\in\real^D$
s.t.~$\signal{S}_{k+1}\widetilde{\signal{z}}^* \neq\signal{S}_k\widetilde{\signal{z}}^*$.
For such a $\widetilde{\signal{z}}^*$, it holds that
$\signal{\Phi}_k\signal{S}_k\widetilde{\signal{z}}^*
=\signal{S}_{k+1}\signal{S}_k^T \signal{S}_k\widetilde{\signal{z}}^*
=\signal{S}_{k+1}\widetilde{\signal{z}}^*
\neq\signal{S}_k\widetilde{\signal{z}}^*$, implying
$\signal{S}_k\widetilde{\signal{z}}^*\not\in\Fix{\signal{\Phi}_k}$.
Hence, we obtain
\begin{eqnarray}
\norm{\signal{\Phi}_k \signal{z}^* - \signal{0}}=
\norm{\signal{S}_{k+1} \widetilde{\signal{z}}^*} = \norm{\signal{S}_k \widetilde{\signal{z}}^*}
=\norm{\signal{z}^* - \signal{0}},
\end{eqnarray}
where $\signal{z}^*:=\signal{S}_k\widetilde{\signal{z}}^*\in
\real^N\setminus \Fix{\signal{\Phi}_k}$
and $\signal{0}\in\Fix{\signal{\Phi}_k}$.
This verifies that $\signal{\Phi}_k$ is not attracting nonexpansive.
\migip

\setcounter{appnum}{3}

\setcounter{equation}{0}
\renewcommand{\theequation}{\Alph{appnum}.\arabic{equation}}

\section{Proof of Theorem \ref{prop:optimality}}
\label{append:prop_optimality}
\noindent{\it Proof of (a)-(I)}:
If $\Theta'_k(\signal{h}_k) = \signal{0}$,
then, $\forall \signal{h}_{(k)}^*\in\Omega_k$,
\begin{eqnarray}
\norm{\signal{h}_{k+1} - \signal{h}_{(k)}^*}^2 &=&
\norm{\signal{\Phi}_k\signal{h}_k - \signal{\Phi}_k\signal{h}_{(k)}^*}^2
\nonumber\\
&\leq& \norm{\signal{h}_k - \signal{h}_{(k)}^*}^2.
\end{eqnarray}
Assume now
$\Theta'_k(\signal{h}_k) \neq \signal{0}$.
In this case, we have
\begin{eqnarray}
&&\norm{\signal{h}_{k+1} - \signal{h}_{(k)}^*}^2 \nonumber\\
&=&
\norm{\signal{\Phi}_k
\left[
\signal{h}_k
-\lambda_k\displaystyle
\frac{\Theta_k(\signal{h}_k)}{\norm{\Theta_k'(\signal{h}_k)}^2}
\Theta'_k(\signal{h}_k) \right]
- \signal{\Phi}_k \signal{h}_{(k)}^*}^2\nonumber\\
&\leq&
\norm{
\signal{h}_k - \signal{h}_{(k)}^*
-\lambda_k\displaystyle
\frac{\Theta_k(\signal{h}_k)}{\norm{\Theta_k'(\signal{h}_k)}^2}
\Theta'_k(\signal{h}_k)}^2\nonumber\\
&=& \norm{\signal{h}_k - \signal{h}_{(k)}^*}^2
-2\lambda_k\displaystyle
\frac{\Theta_k(\signal{h}_k)}{\norm{\Theta_k'(\signal{h}_k)}^2}
\innerprod{\Theta_k'(\signal{h}_k)}
{\signal{h}_k - \signal{h}_{(k)}^*}\nonumber\\
&&
+ \lambda_k^2\displaystyle
\frac{\Theta_k^2(\signal{h}_k)}{\norm{\Theta_k'(\signal{h}_k)}^2} \nonumber\\
&\leq& \norm{\signal{h}_k - \signal{h}_{(k)}^*}^2
- \lambda_k\left[
2\left(1 - \frac{\Theta_k^*}{\Theta_k(\signal{h}_k)}
\right) - \lambda_k
\right]\nonumber\\
&&\hspace*{7em} \times \displaystyle
\frac{\Theta_k^2(\signal{h}_k)}{\norm{\Theta_k'(\signal{h}_k)}^2},
\label{eq:proof_monotonicity}
\end{eqnarray}
which verifies \refeq{eq:mono_appro}.
Here, the first and second inequalities are verified by the nonexpansivity of
$\signal{\Phi}_k$ and the definition of subgradient (see Lemma
\ref{fact:properties} and Appendix \ref{append:mathdef}), respectively.

\noindent{\it Proof of (a)-(II)}: Noting that
$\Theta_k(\signal{h}_k)>\inf_{\signal{x}\in\real^N}\Theta_k(\signal{x})$
implies $\Theta'_k(\signal{h}_k)\neq \signal{0}$,
we can readily verify \refeq{eq:monotone_aII} by \refeq{eq:proof_monotonicity}.

\noindent{\it Proof of (b)}: From Theorem \ref{prop:optimality}.a.I,
we see that
the nonnegative sequence $(\norm{\signal{h}_k - \signal{\omega}})_{k\geq
K_0}$ for any $\signal{\omega}\in \Omega$ is convergent,
hence $(\signal{h}_k)_{k\in \Natural}$ is bounded.
Moreover, since
$\signal{0}\in\partial\Theta_k(\signal{h}_k)$ implies
$\Theta_k(\signal{h}_k)=0$, it is sufficient to check the
case $\Theta'_k(\signal{h}_k)\neq \signal{0}$.
In this case, by \refeq{eq:proof_monotonicity}, we have
\begin{equation}
\norm{\signal{h}_k - \signal{\omega}}^2  -  \norm{\signal{h}_{k+1} -
\signal{\omega}}^2
\geq \varepsilon_1\varepsilon_2 \displaystyle
\frac{\Theta_k^2(\signal{h}_k)}{\norm{\Theta_k'(\signal{h}_k)}^2} \geq 0.
\end{equation}
Therefore, the convergence of
$(\norm{\signal{h}_k - \signal{\omega}})_{k\geq K_0}$
implies
\begin{equation}
\lim_{k\rightarrow
\infty}\frac{\Theta_k^2(\signal{h}_k)}{\norm{\Theta_k'(\signal{h}_k)}^2}=0,
\end{equation}
hence the boundedness of
$(\Theta_k'(\signal{h}_k))_{k\geq \Natural}$ ensures
$\lim_{k\rightarrow \infty, \Theta_k'(\signal{h}_k)\neq \signal{0}}
\Theta_k(\signal{h}_k) = 0$.
\migip


\begin{thebibliography}{100}

\bibitem{tufts}
D. W. Tufts, R. Kumaresan, and I. Kirsteins, "Data adaptive
signal estimation by singular value decomposition of a data matrix,"
Proc. IEEE, vol. 70, pp. 684–685, Jun. 1982.

\bibitem{gabriel}
W. F. Gabriel, "Using spectral estimation techniques in
adaptive processing antenna systems," IEEE Trans. Antennas
Propagat., vol. AP-34, pp. 291–300, Mar. 1986.

\bibitem{scharf}
L. L. Scharf and D. W. Tufts, "Rank reduction for modeling
stationary signals," IEEE Trans. Acoustics, Speech and Signal
Processing, vol. ASSP-35, no. 3, pp. 350–355, Mar. 1987.

\bibitem{vanveen}
B. D. Van Veen and R. A. Roberts, "Partially adaptive
beamformer design via output power minimization," IEEE Trans.
Acoust., Speech, Signal Processing, vol. ASSP-35, pp. 1524–1532,
Nov. 1987.

\bibitem{scharf_svd}
L. L. Scharf, "The SVD and reduced rank signal processing,"
Signal Processing, vol. 25, no. 2, pp. 113–133, 1991.

\bibitem{haimovich}
A. M. Haimovich and Y. Bar-Ness, "An eigenanalysis
interference canceler," IEEE Trans. Signal Processing, vol. 39, no.
1, pp. 76–84, Jan. 1991.

\bibitem{csm}
J. S. Goldstein and I. S. Reed, ``Reduced-rank adaptive
filtering," IEEE Trans. Signal Processing, vol. 45, no. 2, pp.
492–496, Feb. 1997.

\bibitem{wang&poor}
X. Wang and H. V. Poor, ``Blind multiuser detection: A subspace
approach," IEEE Trans. Inform. Theory, vol. 44, no. 2, pp. 677–690,
Mar. 1998.

\bibitem{strom}
E. G. Strom and S. L. Miller, ``Properties of the single-bit
single-user MMSE receiver for DS-CDMA system," IEEE Trans. Commun.,
vol. 47, pp. 416–425, Mar. 1999.

\bibitem{song&roy}
Y. Song and S. Roy, ``Blind adaptive reduced-rank detection for
DSCDMA signals in multipath channels," IEEE J. Selected Areas in
Commun., vol. 17, no. 11, pp. 1960–1970, Nov. 1999.

\bibitem{adapint}
R. C. de Lamare and R. Sampaio-Neto, ``Adaptive reduced-rank MMSE
filtering with interpolated FIR filters and adaptive interpolators,"
IEEE Signal Processing Lett., vol. 12, no. 3, pp. 177–180, Mar.
2005.

\bibitem{jidf_echo}
M. Yukawa, R. C. de Lamare, and R. Sampaio-Neto, "Efficient acoustic
echo cancellation with reduced-rank adaptive filtering based on
selective decimation and adaptive interpolation," IEEE Trans. Audio,
Speech and Language Processing, vol. 56, no. 4, pp. 696–710, May
2008.

\bibitem{jidf_conf}
R. C. de Lamare and R. Sampaio-Neto, ``Adaptive Reduced-Rank MMSE
Parameter Estimation based on an Adaptive Diversity Combined
Decimation and Interpolation Scheme", Proc. IEEE International
Conference on Acoustics, Speech and Signal Processing, April 15-20,
2007, vol. 3, pp. III-1317-III-1320.

\bibitem{st_jidf}
R. C. de Lamare, R. Sampaio-Neto, ``Space–time adaptive reduced-rank
processor for interference mitigation in DS-CDMA systems",
\textit{IET communications}, vol. 2, no. 2, pp. 388-397, 2008.

\bibitem{jidf}
R. C. de Lamare and R. Sampaio-Neto, ``Adaptive Reduced-Rank
Processing Based on Joint and Iterative Interpolation, Decimation,
and Filtering," \textit{IEEE Transactions on Signal Processing},
vol. 57,  no. 7,  July 2009, pp. 2503 - 2514.

\bibitem{barc}
R.C. de Lamare, R. Sampaio-Neto and M. Haardt, "Blind Adaptive
Constrained Constant-Modulus Reduced-Rank Interference Suppression
Algorithms Based on Interpolation and Switched Decimation,"
\textit{IEEE Trans. on Signal Processing},  vol.59, no.2,
pp.681-695, Feb. 2011.

\bibitem{sheng}
S. Li, R. C. de Lamare, R. Fa, ``Reduced-rank linear interference
suppression for DS-UWB systems based on switched approximations of
adaptive basis functions", \textit{IEEE Transactions onVehicular
Technology}, vol. 60, no. 2, pp. 485-497, 2011.

\bibitem{moshavi}
S. Moshavi, E. G. Kanterakis, and D. L. Schilling, ``Multistage
linear receivers for DS-CDMA systems," International Journal of
Wireless Information Networks, vol. 3, no. 1, pp. 1–17, 1996.

\bibitem{mswf}
J. S. Goldstein, I. S. Reed, and L. L. Scharf,``A multistage
representation of the Wiener filter based on orthogonal
projections," IEEE Trans. Signal Processing, vol. 44, no. 7, pp.
2943–2959, Nov. 1998.

\bibitem{honig&xiao}
M. L. Honig and W. Xiao, ``Performance of reduced-rank linear
interference suppression," IEEE Trans. Inform. Theory, vol. 47, no.
5, pp. 1928–1946, July 2001.

\bibitem{honig&goldstein}
M. L. Honig and J. S. Goldstein, ``Adaptive reduced-rank
interference suppression based on multistage Wiener filter," IEEE
Trans. Commun., vol. 50, no. 6, pp. 986–994, Jun. 2002.

\bibitem{mswfccm}
R. C. de Lamare, M. Haardt and R. Sampaio-Neto, "Blind Adaptive
Constrained Reduced-Rank Parameter Estimation based on Constant
Modulus Design for CDMA Interference Suppression," \textit{ IEEE
Transactions on Signal Processing}, vol. 56., no. 6, June 2008.

\bibitem{kansal}
A. Kansal, S. N. Batalama, and D. A. Pados, ``Adaptive maximum SINR
RAKE filtering for DS-CDMA multipath fading channels," IEEE J.
Selected Areas in Commun., vol. 16, no. 9, pp. 1765–1773, Dec. 1998.

\bibitem{pados}
D. A. Pados and S. N. Batalama, ``Joint space-time auxiliary-vector
filtering for DS/CDMA systems with antenna arrays," IEEE Trans.
Communications, vol. 47, no. 9, pp. 1406–1415, Sep. 1999.

\bibitem{honig&xiao}
M. L. Honig and W. Xiao, ``Adaptive reduced-rank interference
suppression with adaptive rank selection," in Proc. Milcom, 2000,
vol. 2, pp. 747–751.




\bibitem{LWF10}
R. C. de Lamare, L. Wang, and R. Fa, ``Adaptive reduced-rank LCMV
beamforming algorithms based on joint iterative optimization of
filters: Design and analysis," \textit{Signal Processing}, vol. 90,
no. 2, pp. 640-652, Feb 2010.

\bibitem{delamarespl07} R. C. de Lamare and R.
Sampaio-Neto, ``Reduced-Rank Adaptive Filtering Based on Joint
Iterative Optimization of Adaptive Filters", \textit{IEEE Signal
Processing Letters}, Vol. 14, no. 12, December 2007.

\bibitem{WLY10}
L Wang, RC de Lamare, M Yukawa, ``Adaptive reduced-rank constrained
constant modulus algorithms based on joint iterative optimization of
filters for beamforming", \textit{IEEE Transactions on Signal
Processing}, vol. 58, no. 6, pp. 2983-2997, 2010.

\bibitem{delamaretvt10}
R. C. de Lamare and R. Sampaio-Neto, ``Reduced-Rank Space-Time
Adaptive Interference Suppression With Joint Iterative Least Squares
Algorithms for Spread-Spectrum Systems," \textit{IEEE Transactions
on Vehicular Technology}, vol.59, no.3, March 2010, pp.1217-1228.

\bibitem{fa_stap}
R. Fa and R. C. de Lamare, ``Reduced-rank STAP algorithms using
joint iterative optimization of filters", \textit{IEEE Transactions
on Aerospace and Electronic Systems}, vol. 47, no. 3, pp. 1668-1684,
2011.


\bibitem{chowdhury}
S. Chowdhury and M. D. Zoltowski, ``Application of conjugate
gradient methods in MMSE equalization for the forward link of
DS-CDMA," in Proc. IEEE VTC 2001-Fall, Oct. 2001, pp. 2434–2438.




\bibitem{Wang}
L. Wang, and R.C.de Lamare ,  ``Constrained adaptive filtering
algorithms based on conjugate gradient techniques for beamforming ",
\textit{IET Signal Processing}, vol. 4, issue. 6, pp. 686-697, Feb.
2010.

\bibitem{FLW10} R. Fa, R. C. de
Lamare and L. Wang, ``Reduced-rank STAP schemes for airborne radar
based on switched joint interpolation, decimation and filtering
algorithm", \textit{IEEE Trans. Sig. Proc.}, 2010, vol. 58, no. 8,
pp.4182-4194.

\bibitem{chen} W. Chen, U. Mitra, and P. Schniter, ``On
the equivalence of three reduced rank linear estimators with
applications to DS-CDMA," IEEE Trans. Inform. Theory, vol. 48, no.
9, pp. 2609-2614, Sept. 2002.

\bibitem{burykh}
S. Burykh and K. Abed-Meraim, "Reduced-rank adaptive filtering using
Krylov subspace," EURASIP J. Appl. Signal Processing, no. 12, pp.
1387-1400, Dec. 2002.

\bibitem{dietl}
G. K. E. Dietl, Linear estimation and detection in Krylov
subspaces —Foundations in signal processing, communications and
networking, Springer, 2007.

\bibitem{gollamudi}
S. Gollamudi, S. Nagaraj, S. Kapoor, and Y. H. Huang,
``Set-membership filtering and a set-membership normalized LMS
algorithm with an adaptive step size," IEEE Signal Processing Lett.,
vol. 5, no. 5, pp. 111–114, May 1998.

\bibitem{guo}
L. Guo, A. Ekpenyong, and Y. H. Huang, ``Frequency-domain adaptive
filtering —A set-membership approach," in Proc. Asilomar Conf.
Signals, Syst., Comput., 2003, pp. 2073-2077.

\bibitem{yamada}
I. Yamada, K. Slavakis, and K. Yamada, ``An efficient robust
adaptive filtering algorithm based on parallel subgradient
projection techniques," IEEE Trans. Signal Processing, vol. 50, no.
5, pp. 1091–1101, May 2002.

\bibitem{delamaresmf}
R. C. de Lamare and P. S. R. Diniz, ``Set-Membership Adaptive
Algorithms based on Time-Varying Error Bounds for CDMA Interference
Suppression", \emph{IEEE Trans. on Vehicular Technology}, vol. 58,
no. 2,  February 2009 , pp. 644 - 654. \vspace{-0.1em}

\bibitem{clarke_smjio}
P. Clarke and R. C. de Lamare, "Low-Complexity Reduced-Rank Linear
Interference Suppression based on Set-Membership Joint Iterative
Optimization for DS-CDMA Systems",  IEEE Trans. on Vehicular
Technology, vol. 60, no. 9, 2011, pp. 4324-4337.

\bibitem{ce_wang}
T. Wang, R. C. de Lamare and P. D. Mitchell, ``Low-complexity
set-membership channel estimation for cooperative wireless sensor
networks", IEEE Transactions on Vehicular Technology, vol. 60, no.
6, pp. 2594-2607, 2011.






\bibitem{yukawa}
M. Yukawa and I. Yamada, ``Pairwise optimal weight realization —
Acceleration technique for set-theoretic adaptive parallel
subgradient projection algorithm," IEEE Trans. Signal Processing,
vol. 54, no. 12, pp. 4557–4571, Dec. 2006.



\bibitem{yukawa2}
M. Yukawa, K. Slavakis, and I. Yamada,``Adaptive parallel
quadraticmetric projection algorithms," IEEE Trans. Audio, Speech
and Language Processing, vol. 15, no. 5, pp. 1665–1680, July 2007.

%
%
%

\bibitem{haykin}
S. Haykin, Adaptive Filter Theory, Prentice Hall, New Jersey,
4th edition, 2002.

\bibitem{golub}
G. H. Golub and C. F. Van Loan, Matrix Computations, The Johns
Hopkins University Press, Baltimore, 3rd edition, 1996.

\bibitem{horn}
R. A. Horn and C. R. Johnson, Matrix Analysis, Cambridge
University Press, Cambridge, 1985.

\bibitem{axelsson}
O. Axelsson, ``A survey of preconditioned iterative methods for
linear systems of algebraic equations," BIT, vol. 25, pp. 166–187,
1985.

\bibitem{saad}
Y. Saad, Iterative methods for sparse linear systems, SIAM,
Philadelphia, PA, 2nd edition, 2003.

\bibitem{hull}
A. W. Hull and W. K. Jenkins, ``Preconditioned conjugate gradient
methods for adaptive filtering," in Proc. IEEE Int. Symp. Circuits
Syst., Jun. 1991, pp. 540–543.

\bibitem{boray}
G. K. Boray and M. D. Srinath, ``Conjugate gradient techniques for
adaptive filtering," IEEE Trans. Circuits Syst. I, vol. 39, no. 1,
pp. 1–10, Jan. 1992.

\bibitem{chang}
P. S. Chang and A. N. Willson, Jr., ``Analysis of conjugate gradient
algorithms for adaptive filtering," IEEE Trans. Signal Processing,
vol. 48, no. 2, pp. 409–418, Feb. 2000.


\bibitem{hinamoto}
T. Hinamoto and S. Maekawa, ``Extended theory of learning
identification," Trans. IEE Japan, vol. 95, no. 10, pp. 227–234,
1975, in Japanese.

\bibitem{ozeki}
K. Ozeki and T. Umeda, ``An adaptive filtering algorithm using an
orthogonal projection to an affine subspace and its properties,"
IEICE Trans., vol. 67-A, no. 5, pp. 126–132, 1984, in Japanese.


\bibitem{combettes}
P. L. Combettes, ``The foundations of set theoretic estimation,"
Proc. IEEE, vol. 81, no. 2, pp. 182–208, Feb. 1993.

\bibitem{bauschke}
H. H. Bauschke and J. M. Borwein, ``On projection algorithms for
solving convex feasibility problems," SIAM Review, vol. 38, no. 3,
pp. 367–426, 1996.

\bibitem{censor}
Y. Censor and S. A. Zenios, Parallel Optimization: Theory,
Algorithm, and Optimization, Oxford University Press, 1997.


\bibitem{bunariu}
D. Butnariu, Y. Censor, and S. Reich, Eds., Inherently parallel
algorithms in feasibility and optimization and their applications,
New York: Elsevier, 2001.


\bibitem{madhow}
U. Madhow and M. L. Honig, ``MMSE interference suppression for
direct-sequence spread-spectrum CDMA," IEEE Trans. Commun., vol. 42,
no. 12, pp. 3178–3188, Dec. 1994.

\bibitem{boyd}
S. Boyd and L.Vandenberghe, Convex optimization, Cambridge
University Press, Cambridge, 2004.

\bibitem{strang}
G. Strang, Linear algebra and its applications, Saunders
College Publishing, 3rd edition, 1988.

\end{thebibliography}
%

\end{document}